\documentclass[%
 reprint,
superscriptaddress,
nofootinbib,
 amsmath,amssymb,
 aps,
 prd, 
 showkeys,
 nolongbibliography,
]{revtex4-2}

\usepackage{graphicx}
\usepackage{dcolumn}
\usepackage{bm}
\usepackage{multirow}
\usepackage{xcolor}
\usepackage[
    colorlinks=true,
    linkcolor=red,        
    citecolor=blue,       
    urlcolor=magenta      
]{hyperref}
\usepackage[inline]{enumitem}
\usepackage{booktabs}
\usepackage{soul}
\usepackage{amsmath, amssymb, amsfonts}
\usepackage{xcolor}

\usepackage{xcolor}
\usepackage{tikz}
\usepackage[percent]{overpic}
\definecolor{mygray}{RGB}{120,120,120}
\definecolor{dodgerblue}{RGB}{30,144,255}
\definecolor{tomato}{RGB}{255,99,71}

\begin{document}

\title{Revisiting Metastable Dark Energy in Light of DESI DR2 BAO and DESI DR1 Full-Shape Measurements}

\author{Purba Mukherjee}
\email{purba@kasi.re.kr}
\affiliation{Korea Astronomy and Space Science Institute, 776 Daedeok-daero, Yuseong-gu, Daejeon 34055, Republic of Korea}%
\author{Arman Shafieloo}%
\email{shafieloo@kasi.re.kr}
\affiliation{Korea Astronomy and Space Science Institute, 776 Daedeok-daero, Yuseong-gu, Daejeon 34055, Republic of Korea}%
\affiliation{University of Science and Technology, Daejeon 34113, Korea}
\author{Varun Sahni}%
\email{varun@iucaa.in}
\affiliation{Inter-University Centre for Astronomy and Astrophysics, Post Bag 4, Ganeshkhind, Pune 411007, India}%


\date{\today}

\begin{abstract}
We revisit metastable dark energy (DE) models described by a radioactive-like decay law. We consider three scenarios: an effective, exponentially decaying DE component; decay of DE into non-baryonic dark matter (DM); and decay of DE into dark radiation (DR).  We constrain the metastable DE models using DESI DR2 baryon acoustic oscillation (BAO) data, Type Ia supernovae (SNIa), cosmic microwave background (CMB) observations, and, for the first time, the current available DESI DR1 full-shape (FS) clustering measurements.  The BAO+SNIa combinations show a mild preference for positive $\Gamma/H_0$, with a deviation from the $\Lambda$CDM limit at the $\gtrsim 2\sigma$ level. This corresponds to a decaying DE density and an effective quintessence-like behaviour at low redshift. Once CMB information from either Planck or P-ACT is included, however, the constraints become statistically consistent with $\Gamma/H_0=0$. The FS measurements probe the growth sector and help distinguish the interacting DM-DE behaviour of Model 2 from the effective decaying-DE response of Model 1 and the weaker DR-induced response of Model 3. For CMB+FS+DES-Dovekie, Model 1 shows a slight deviation from $\Gamma/H_0=0$ at the $\gtrsim 2\sigma$ level, while Models 2 and 3 remain consistent with the $\Lambda$CDM limit within $2\sigma$. Overall, metastable DE remains phenomenologically viable: current data allow late-time dynamics but do not provide decisive evidence for a nonzero decay rate. These results motivate extending our analysis to the upcoming DESI DR2 FS data to obtain tighter constraints on metastable dynamics.

\end{abstract}

\keywords{cosmology, dark energy, growth rate of structure, Hubble tension}
\maketitle
\section{Introduction}
\label{sec:intro}

The standard $\Lambda$CDM model has long served as the baseline framework for interpreting cosmological observations, including the cosmic microwave background (CMB) \cite{Planck:2018vyg, AtacamaCosmologyTelescope:2025blo, SPT-3G:2025bzu}, Type Ia supernovae (SN-Ia) \cite{Scolnic:2021amr, DES:2024jxu}, baryon acoustic oscillation (BAO) and large-scale structure (LSS) surveys \cite{eBOSS:2020yzd, DESI:2025zgx, DESI:2024jxi}. In this picture, the matter sector is dominated by cold, pressureless dark matter, while cosmic acceleration is attributed to a cosmological constant, $\Lambda$, or equivalently to a dark energy (DE) component with equation of state $w=-1$ \cite{Einstein:1917ce, Zeldovich:1968ehl}. However, the nature of DE remains elusive, and the cosmological constant problem remains a fundamental theoretical challenge \cite{Weinberg:1988cp, Padmanabhan:2002ji}. At the same time, increasingly precise cosmological data have revealed persistent tensions between different probes \cite{DiValentino:2025otz}. These tensions suggest that the dark sector may be more complex than in the minimal $\Lambda$CDM scenario \cite{CosmoVerseNetwork:2025alb}. One of the most widely discussed discrepancies is the $H_0$ tension \cite{Verde:2019ivm}. Local measurements from the $H_0$ Distance Network (H0DN) report a consensus value of $H_0$ that is in $\sim7\sigma$ tension with early-Universe $\Lambda$CDM constraints based on SPT-3G, Planck, and ACT \cite{H0DN:2025lyy}. Growth-sector constraints, commonly summarized by $S_8$, also show mild but persistent deviations between CMB and LSS measurements \cite{DES:2026fyc}. At higher redshifts, JWST observations of massive and luminous galaxy candidates have raised questions about early structure formation and the efficiency of galaxy assembly within the standard cosmological framework \cite{Labbe:2022ahb, Boylan-Kolchin:2022kae}. While uncertainties remain, these results further motivate testing extensions to the baseline $\Lambda$CDM model.

Recent DESI BAO measurements provide tantalizing hints for dynamical dark energy \cite{DESI:2025zgx}, with CMB+SN-Ia combinations showing $\gtrsim 3\sigma$ deviations from $w=-1$ within the Chevallier-Polarski-Linder \cite{Chevallier:2000qy, Linder:2002et} (CPL) parametrization, depending on the data set employed. Similar indications have also been found in model-independent reconstructions, where the effective DE equation of state (EoS) or pressure can show deviations from $\Lambda$CDM \cite{DESI:2025fii, Mukherjee:2025ytj, Berti:2025phi, Ormondroyd:2025exu, DES:2025sig}. Interestingly, these reconstructions often point to a phantom-crossing behaviour \cite{Caldwell:1999ew}, where the DE EoS crosses $w=-1$. Such a crossing is difficult to realize in simple single-field DE models \cite{Ratra:1987rm, Caldwell:1997ii} and may require more exotic dark-sector physics, modified gravity, or a different microscopic description of DE. This motivates testing alternative scenarios in which the apparent evolution of the DE sector arises from different physical mechanisms (see Refs. \cite{Khoury:2025txd, Giare:2024smz, Shah:2025ayl, Chakraborty:2025syu, Mukherjee:2025myk, Wolf:2025acj, Naidoo:2026umv, Mishra:2025goj, Guedezounme:2025wav, Li:2026hwq, Gomez-Valent:2026ept, Parker:1968mv, Landim:2016isc, Jiang:2026cqh}). One such possibility is metastable dark energy, in which the DE density evolves through a radioactive-like decay law \cite{Shafieloo:2016bpk}.

In this framework, the DE density is not strictly constant, but evolves according to a radioactive-like decay law \cite{Shafieloo:2016bpk}. The decay rate is treated as an intrinsic property of the DE sector, rather than being fixed by the Hubble expansion rate or curvature. Previous studies have shown that  Metastable models can provide a viable description of an evolving DE sector that can be tested against observations \cite{Li:2019san,Yang:2020zuk}. Current DESI observations therefore provide a timely opportunity to revisit metastable models using BAO and full-shape clustering measurements, together with SNIa and CMB data. Here, we consider three metastable DE scenarios. In Model 1, the DE density decays effectively as a background component, leading to a redshift-dependent effective equation of state. In Model 2, DE decays into non-baryonic dark matter (DM), behaving as an interacting DM-DE scenario that affects both the matter abundance and perturbations.  In Model 3, DE decays into dark radiation (DR), producing an additional relativistic component {at late times}. These three scenarios help distinguish modifications confined to DE only from those that affect the matter or radiation sectors.
We constrain {Metastable} models using DESI DR2 BAO \cite{DESI:2025fii} measurements, SN-Ia compilations (DES-Dovekie \cite{DES:2025sig}, Union3 \cite{Rubin:2023jdq}, PantheonPlus \cite{Brout:2022vxf}), a BBN prior for physical baryon density \cite{Schoneberg:2024ifp}, and CMB observations from Planck \cite{Rosenberg:2022sdy, Planck:2019nip} and Planck+ACT, denoted P-ACT \cite{AtacamaCosmologyTelescope:2025blo}. 
We extend earlier background-level studies by consistently evolving linear perturbations. Energy transfer from DE to DM modifies the DM density perturbations, while the DR model requires the evolution of the relativistic perturbation hierarchy. This framework connects dark sector microphysics not only to cosmic distances but also to structure formation through the power spectrum. 

A key novelty of our work is the first application of DESI DR1 full-shape measurements \cite{DESI:2024jxi} to metastable DE models, extending the analysis beyond BAO distances to include growth and clustering information. Since the three decay channels affect the matter sector and radiation sector differently, full-shape data help test their distinct signatures and their consistency with structure growth. Our work therefore demonstrates the unique discriminatory power of full-shape measurements. This joint treatment of expansion, perturbations, and scale-dependent clustering represents the future of redshift-survey cosmology. 

The paper is organized as follows. In Sec.~\ref{sec:model}, we describe the three metastable DE models and their background and perturbation equations. In Sec.~\ref{sec:data}, we summarize the data sets, priors, and analysis methodology. In Sec.~\ref{sec:results}, we present the parameter constraints from diverse data combinations. In Sec.~\ref{sec:evo}, we discuss reconstructed background and growth observables. In Sec.~\ref{sec:fs}, we study the impact of DESI DR1 full-shape measurements. In Sec.~\ref{sec:model_comparison}, we compare the relative goodness of fit of the metastable DE models with respect to $\Lambda$CDM. Finally, we summarize our conclusions in Sec.~\ref{sec:summary}.

\section{Formalism \label{sec:model}}

We consider a spatially flat Friedmann-Lema\^{i}tre-Robertson-Walker (FLRW) Universe,
\begin{equation}
    \mathrm{d}s^2 = c^2 \mathrm{d}t^2 
    - a^2(t)\left[\mathrm{d}r^2 
    + r^2 \mathrm{d}\theta^2 
    + r^2\sin^2\theta \mathrm{d}\phi^2\right]\, ,
\end{equation}
where $a(t)$ is the scale factor and $c$ is the speed of light. The Hubble expansion rate is given by
\begin{equation}
    \frac{H^2(z)}{H_0^2} = 
    \Omega_{m}(1+z)^3
    + \Omega_{r}(1+z)^4
    + \Omega_{\rm DE} \, ,
\end{equation}
where $H_0$ is the present-day Hubble parameter, $z$ is the redshift, and $\Omega_{m}$, $\Omega_{r}$, and $\Omega_{\rm DE}$ denote the {current values of the} \textit{matter}, \textit{radiation}, and \textit{dark energy} density parameters, respectively. The matter sector consists of \textit{baryons} and \textit{cold dark matter} (CDM), i.e. $\Omega_{m}=\Omega_{b}+\Omega_{c}$ where $\Omega_{b}$ and $\Omega_{c}$ denote the present-day baryon and CDM density parameters. The transverse comoving distance to an object at redshift $z$ is
\begin{equation}
    D_M(z)={c}\int_0^z \frac{\mathrm{d}z'}{H(z')} \, .
\end{equation}

We study three phenomenological metastable dark-energy scenarios: an exponentially decaying DE density (Model 1), DE decay into dark matter (Model 2), and DE decay into dark radiation (Model 3). These cases respectively produce an effective redshift-dependent DE EoS, modify the matter-sector evolution, and add an extra-relativistic component to the energy budget. We assume that the DE EoS approaches $w=-1$ at high redshift, and that its decay arises from intrinsic properties of the DE sector rather than from the expansion history of the Universe. Together, these scenarios allow us to assess the cosmological consequences of a metastable dark sector.

\begin{figure*}
    \centering
    \includegraphics[width=\linewidth, height=0.25\textwidth]{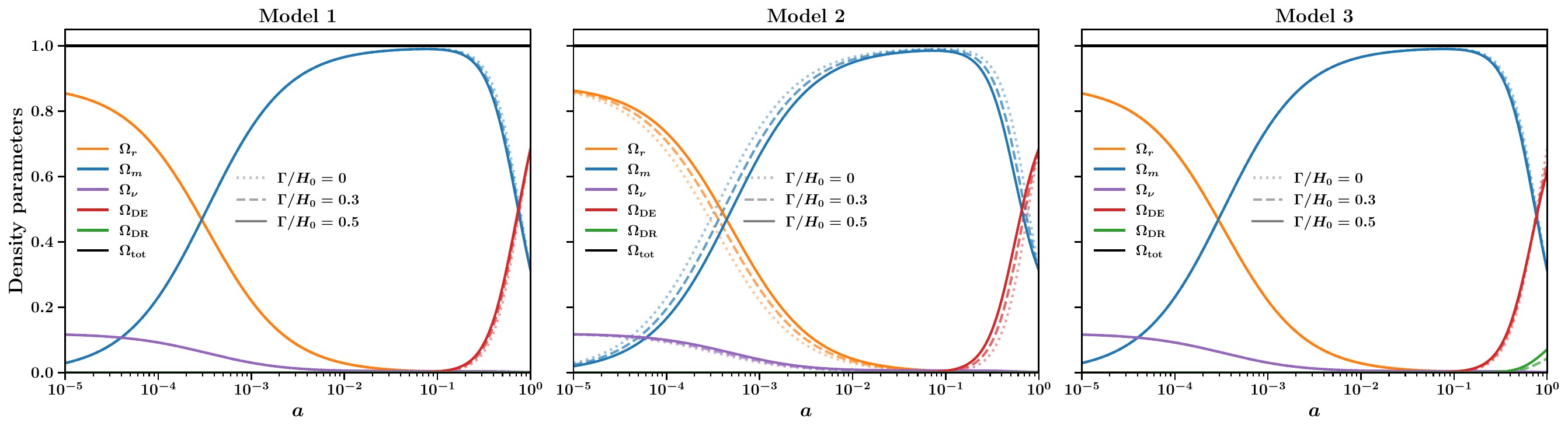}\\
    \includegraphics[width=\linewidth, height=0.25\textwidth]{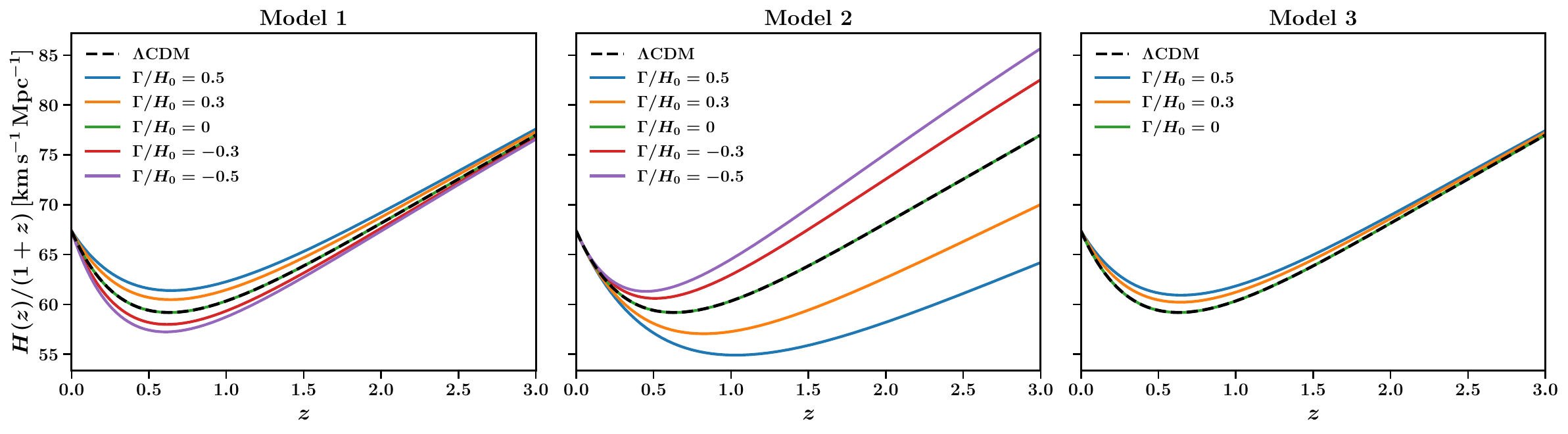}
\caption{Evolution of the fractional density parameters $\Omega_i(a)$ and the Hubble parameter $\frac{H(z)}{1+z}$ for the metastable DE models, different values of $\Gamma/H_0$. The nonzero $\Omega_{\rm DR}$ contribution in Model 3 highlights the DR component produced by DE decay.}
\label{fig:omega_evolution}
\end{figure*}

\subsection{Model 1: Exponentially decaying DE}

The DE sector is modeled as a single effective fluid with a radioactive-decay-like density evolution,
\begin{equation}
    \rho_{\rm DE}(t)=\rho_{\rm DE}(t_0)
    \exp[-\Gamma(t-t_0)] \, ,
\end{equation}
which follows from
\begin{equation}
    \dot{\rho}_{\rm DE}=-\Gamma \rho_{\rm DE} \, .
\end{equation}
Here $\Gamma$ is a constant decay rate with dimensions of inverse time. For $\Gamma>0$, the DE density decreases with time, with half-life $t_{1/2}=\ln 2/\Gamma$, whereas $\Gamma<0$ corresponds to a growing DE density and hence phantom-like behaviour. The DE density parameter evolves as
\begin{equation}
    \Omega_{\rm DE}(z)=\Omega_{{\rm DE},0}
    \exp\left[
    \int_0^z \frac{\Gamma}{H(z'){(1+z')}}{\mathrm{d}z'}
    \right] \, .
\end{equation}
The corresponding Hubble expansion rate is
\begin{equation}
    \frac{H^2(z)}{H_0^2} =
    \Omega_{\rm DE}(z)
    + \Omega_{m}(1+z)^3
    + \Omega_{r}(1+z)^4 \, .
\end{equation} The effective DE pressure follows from the continuity Eq.,
\begin{equation}
    \dot{\rho}_{\rm DE}
    +3H\left(\rho_{\rm DE}+p_{\rm DE}\right)=0 \, ,
\end{equation}
which gives
\begin{equation}
    p_{\rm DE}
    =
    -\rho_{\rm DE}\left(1-\frac{\Gamma}{3H}\right) \, .
\end{equation}
Therefore, the effective DE EoS is
\begin{equation} \label{eq:weff_de}
    w_{\rm DE}(z)
    \equiv
    \frac{p_{\rm DE}}{\rho_{\rm DE}}
    =
    -1+\frac{\Gamma}{3H(z)} \, .
\end{equation}
Thus, the model is governed by the dimensionless parameter $\Gamma/H_0$, which sets the DE decay timescale relative to the Hubble time.

\begin{figure*}
    \centering
    \includegraphics[width=\linewidth, height=0.25\textwidth]{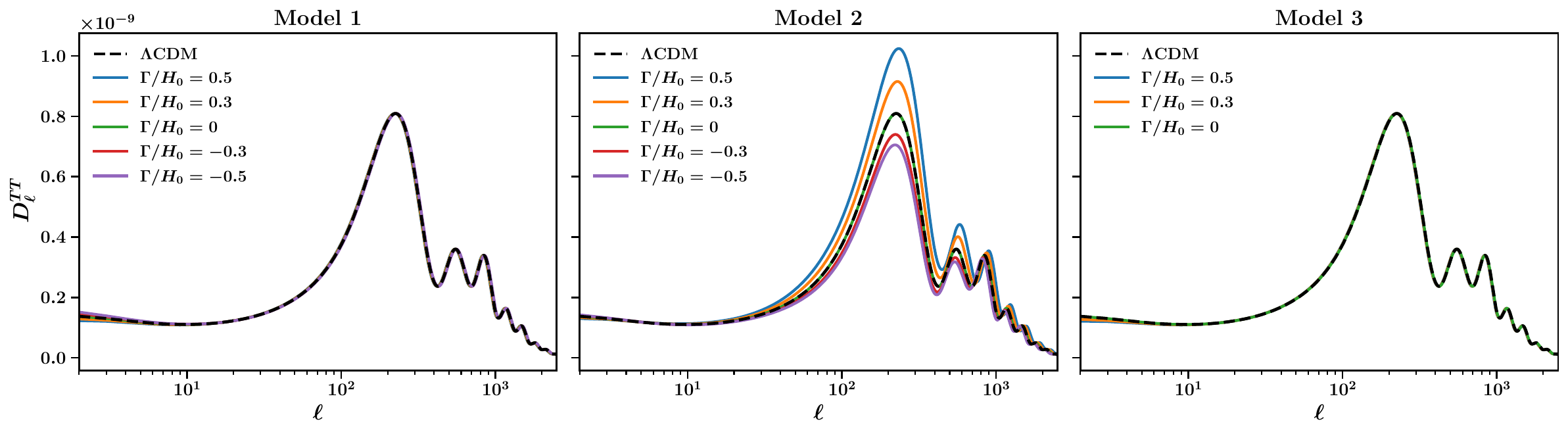}
    \includegraphics[width=\linewidth, height=0.25\textwidth]{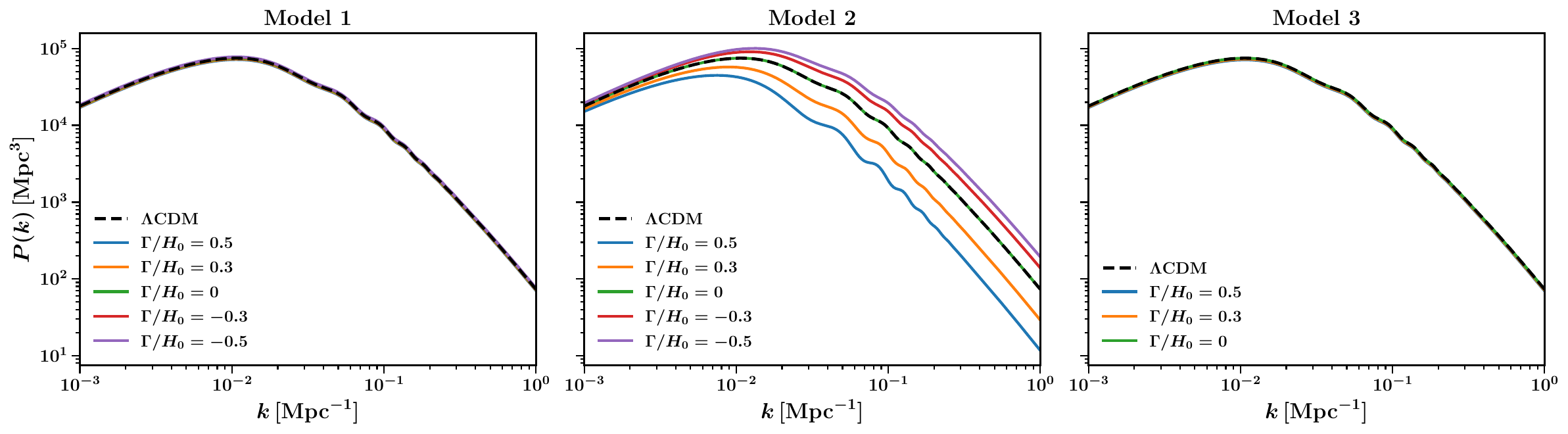}
    \includegraphics[width=\linewidth, height=0.25\textwidth]{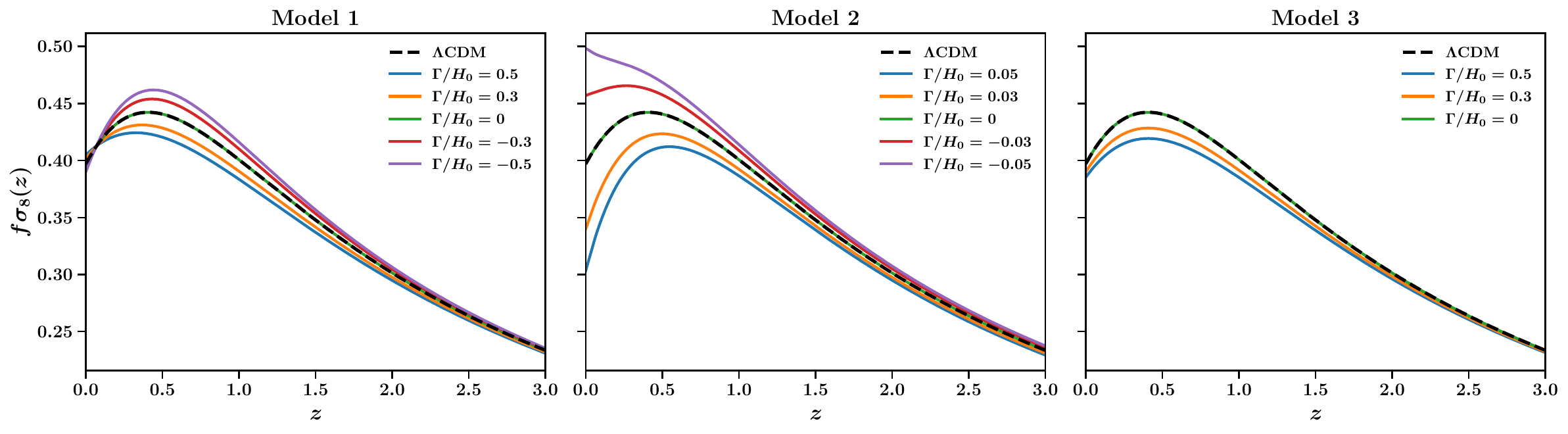}
\caption{
Theoretical evolution of angular power spectrum $D_\ell^{TT}$, matter power spectrum $P(k)$, and growth rate of structure $f\sigma_8(z)$ for the metastable DE models with different values of $\Gamma/H_0$. The panels illustrate the impact of DE decay on the CMB anisotropies and structure growth.}
\label{fig:theory_observables2}
\end{figure*}

\subsection{Model 2: DE decay into DM}

We consider a scenario in which DE decays into non-baryonic DM, while the baryonic sector remains separately conserved. The background evolution is set by
\begin{align}
    \dot{\rho}_{\rm DE} &= -\Gamma \rho_{\rm DE} \, , \nonumber \\
    \dot{\rho}_{\rm DM} + 3H\rho_{\rm DM} &= \Gamma \rho_{\rm DE} \, ,
\end{align}
where $\Gamma$ is a constant decay rate. Prior to its decay, the DE component is assumed to behave as a cosmological constant, i.e., $w=-1$. For $\Gamma \neq 0$, this is effectively an interacting DM-DE model, since energy is exchanged between DM and DE. For $\Gamma>0$, energy is transferred from DE to DM, while $\Gamma<0$ corresponds to energy flow in the opposite direction.

Since only the non-baryonic dark matter component participates in the interaction, the baryon density evolves independently as $\rho_b \propto (1+z)^3$. The Hubble expansion rate is therefore
\begin{equation}
    \frac{H^2(z)}{H_0^2}
=   \Omega_{\rm DE}(z)
  + \Omega_{\rm DM}(z) + \Omega_{b}(1+z)^3
    + \Omega_{r}(1+z)^4 \, .
\end{equation}
At the {perturbative} level, we assume that the DE component remains vacuum-like and does not cluster. The decay affects the growth of structure through the modified DM perturbation equations. In synchronous gauge, with the energy-transfer four-vector parallel to the DM four-velocity, the linear perturbations obey
\begin{align}
    \delta_{\rm DM}' &=
    -\theta_{\rm DM}
    - \frac{h'}{2}
    - a\Gamma
    \frac{\rho_{\rm DE}}{\rho_{\rm DM}}
    \delta_{\rm DM} \, , \nonumber \\
    \theta_{\rm DM}' &=
    -\mathcal{H}\theta_{\rm DM} \, .
\end{align}
Here primes denote derivatives with respect to conformal time, $\mathcal{H}=aH$, $h$ is the trace metric perturbation in synchronous gauge, and $\delta_{\rm DM}\equiv \delta\rho_{\rm DM}/\rho_{\rm DM}$. In the DM-comoving synchronous gauge, setting $\theta_{\rm DM}=0$ gives
\begin{equation}
    \delta_{\rm DM}'
    =
    -\frac{h'}{2}
    - a\Gamma
    \frac{\rho_{\rm DE}}{\rho_{\rm DM}}
    \delta_{\rm DM} \, .
\end{equation}
Thus, $\Gamma$ modifies structure formation through both the background DM abundance and evolution of DM density perturbations.

\subsection{Model 3: DE decay into DR}

We also consider a scenario in which DE decays into dark radiation (DR). Here, the DR component has no primordial abundance and is produced solely through the decay of DE. Therefore, we restrict this scenario to $\Gamma>0$. The background evolution is governed by
\begin{align}
    \dot{\rho}_{\rm DE} &= -\Gamma \rho_{\rm DE} \, , \nonumber \\
    \dot{\rho}_{\rm DR} + 4H\rho_{\rm DR} &= \Gamma \rho_{\rm DE} \, ,
\end{align}
where the produced DR behaves as an ultra-relativistic component with $w_{\rm DR}=1/3$. The non-relativistic matter sector, including baryons and cold dark matter, remains separately conserved, with $\rho_m\propto(1+z)^3$. The Hubble expansion rate is therefore 
\begin{equation}
    \frac{H^2(z)}{H_0^2}
=    \Omega_{\rm DE}(z)
    + \Omega_{\rm DR}(z)
    + \Omega_{m}(1+z)^3
    + \Omega_{r}(1+z)^4 \, , 
\end{equation}
where $\Omega_{m}=\Omega_{b}+\Omega_{c}$, and $\Omega_{r}$ denotes the standard radiation density parameter.

At the perturbative level, the metastable decay sources perturbations in the DR sector. In synchronous gauge, the perturbation equations for the normalized DR density contrast, velocity divergence, and anisotropic stress are
\begin{align}
    \delta_{\rm DR}' &=
    -\frac{4}{3}\theta_{\rm DR}
    -\frac{2}{3}h'
    -a\Gamma
    \frac{\rho_{\rm DE}}{\rho_{\rm DR}}
    \delta_{\rm DR} \, , \nonumber \\
    \theta_{\rm DR}' &=
    k^2
    \left(
    \frac{1}{4}\delta_{\rm DR}
    -\sigma_{\rm DR}
    \right)
    -a\Gamma
    \frac{\rho_{\rm DE}}{\rho_{\rm DR}}
    \theta_{\rm DR} \, , \\
    \sigma_{\rm DR}' &=
    \frac{4}{15}\theta_{\rm DR}
    -\frac{3}{10}kF_{{\rm DR},3}
    -\frac{2}{15}h'
    -\frac{4}{5}\eta'
    -a\Gamma
    \frac{\rho_{\rm DE}}{\rho_{\rm DR}}
    \sigma_{\rm DR} \, . \nonumber
\end{align}
Here primes denote conformal-time derivatives, $h$ and $\eta$ are synchronous-gauge metric perturbations, $\theta_{\rm DR}$ is the DR velocity divergence, $\sigma_{\rm DR}$ is the DR anisotropic stress, and $F_{{\rm DR},3}$ is the octupole moment of DR hierarchy. For higher multipoles, the hierarchy reads
\begin{equation}
\begin{split}
\small    F_{{\rm DR},\ell}'
    =
    \frac{k}{2\ell+1}
    \left[
    \ell F_{{\rm DR},\ell-1}
    -(\ell+1)F_{{\rm DR},\ell+1}
    \right] + \\
    \quad - a\Gamma
    \frac{\rho_{\rm DE}}{\rho_{\rm DR}}
    F_{{\rm DR},\ell}\,\,  \quad \quad
    \qquad \ell \geq 3 \,  ,
\end{split}
\end{equation}
with the usual closure relation imposed at $\ell_{\rm max}$. Thus, in Model 3, DE decay modifies cosmological evolution through both the background production of DR and the perturbations of the relativistic daughter component.

\section{Data and Analysis \label{sec:data}}

To constrain the three metastable DE models, we use several combinations of cosmological data sets. These include the DESI DR2 baryon acoustic oscillation (BAO) measurements \cite{DESI:2025zgx}, Type Ia supernovae (SN-Ia) compilations \cite{Scolnic:2021amr, DES:2025sig, Rubin:2023jdq}, cosmic microwave background (CMB) \cite{Planck:2018vyg, AtacamaCosmologyTelescope:2025blo} data, and a Big Bang nucleosynthesis (BBN) \cite{Schoneberg:2024ifp} prior on the physical baryon density $\omega_b \equiv \Omega_{b} h^2$. For the SNe Ia data, we consider three independent compilations\footnote{We report the DES-Dovekie results in the main text, while the Union3 and PantheonPlus results are given in Appendix~\ref{app:snia}.}, namely DES-Dovekie \cite{DES:2025sig}, Union3 \cite{Rubin:2023jdq}, and PantheonPlus \cite{Brout:2022vxf}. For the CMB sector, we use either the Planck PR4 \texttt{CamSpec NPIPE} TTTEEE spectra \cite{Rosenberg:2022sdy}, supplemented by the PR3 low-$l$ TT and EE likelihoods \cite{Planck:2019nip}; or the Planck+ACT DR6 likelihood \cite{AtacamaCosmologyTelescope:2025blo}, hereafter denoted as P-ACT. The BBN prior is taken from the physical baryon density estimate obtained with the \texttt{PRyMordial} \cite{Burns:2023sgx} code, including marginalisation over uncertainties in nuclear reaction rates \cite{Cooke:2017cwo, Aver_2021}, which gives
\begin{equation}
    \omega_b = 0.02218 \pm 0.00055 \, . \label{eq:bbn}
\end{equation}
We analyse the following data combinations:
\begin{enumerate*}
    \item BBN + DESI,
    \item BBN + DESI + SN-Ia,
    \item CMB + DESI,
    \item CMB + DESI + SN-Ia.
\end{enumerate*}
This allows us to test available CMB and SN-Ia data combinations.

The theoretical predictions are computed using a modified version of \texttt{CLASS}\footnote{\url{https://github.com/lesgourg/class_public.git}} \cite{Lesgourgues:2011re, Blas:2011rf}, which implements the background and perturbation equations for the metastable DE models. Figure~\ref{fig:omega_evolution} shows the evolution of the fractional density parameters and $H(z)/(1+z)$, illustrating how $\Gamma/H_0$ modifies the cosmic energy budget and background expansion history. The effect on the CMB temperature power spectrum $D_\ell^{TT}$, the linear matter power spectrum $P(k)$, and the growth rate of structure $f\sigma_8(z)$ are shown in Figure~\ref{fig:theory_observables2}.

\begin{table}[t]
\centering
\caption{Priors adopted for model parameters.}
\label{tab:priors}
\renewcommand{\arraystretch}{1.2}
\setlength{\tabcolsep}{15pt}
\begin{tabular}{c c}
\toprule
\textbf{Parameter} & \textbf{Prior} \\
\hline
${\boldmath \omega_b}$ & $\mathcal{U}(0.005,\,0.04)$ \\
${\boldmath \omega_c}$ & $\mathcal{U}(0.001,\,0.99)$ \\
${\boldmath H_0}$ & $\mathcal{U}(20,\,100)$ \\
${\boldmath \tau_{\rm reio}}$ & $\mathcal{U}(0.01,\,0.8)$ \\
${\boldmath n_s}$ & $\mathcal{U}(0.8,\,1.2)$ \\
${\boldmath\ln(10^{10}A_s)}$ & $\mathcal{U}(1.61,\,3.91)$ \\
\hline
${\boldmath \Gamma/H_0}$ & $\mathcal{U}(-1,\,2)$ \quad Models 1 and 2 \\
${\boldmath \Gamma/H_0}$ & $\mathcal{U}(0,\,2)$ \quad Model 3 \\
\bottomrule
\end{tabular}
\end{table}

\begin{figure*}
    \centering
    \includegraphics[height=0.22\textheight, width=0.325\linewidth]{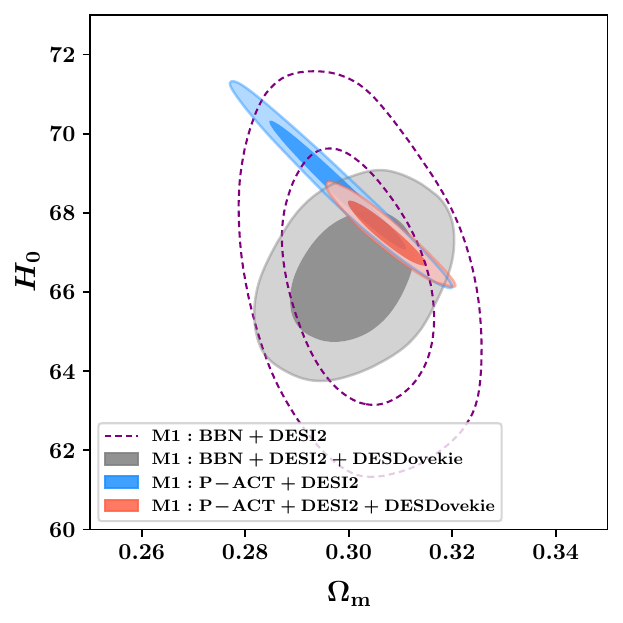}
    \includegraphics[height=0.22\textheight, width=0.325\linewidth]{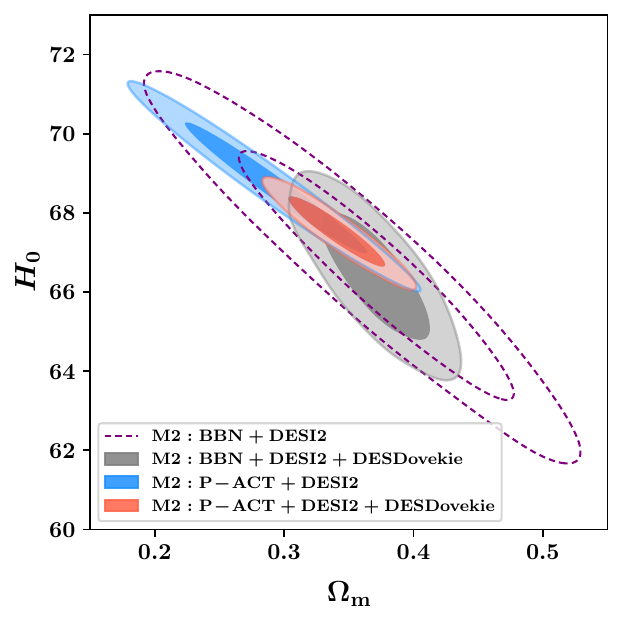}
    \includegraphics[height=0.22\textheight, width=0.325\linewidth]{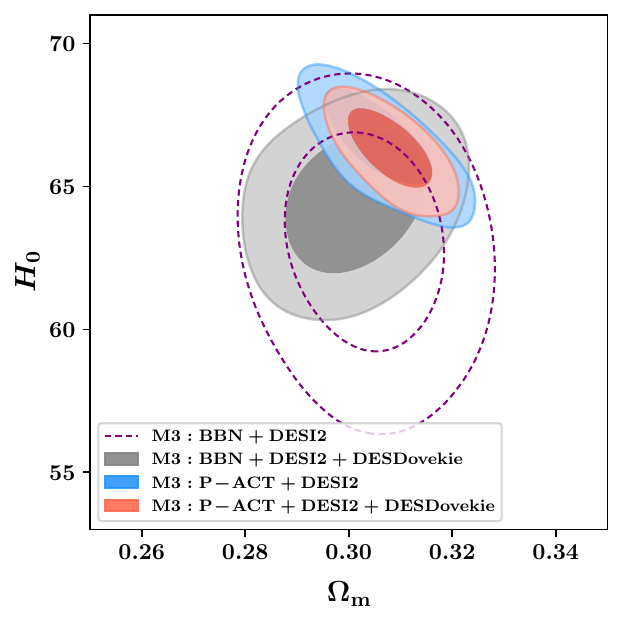}\\
    \includegraphics[height=0.22\textheight, width=0.325\linewidth]{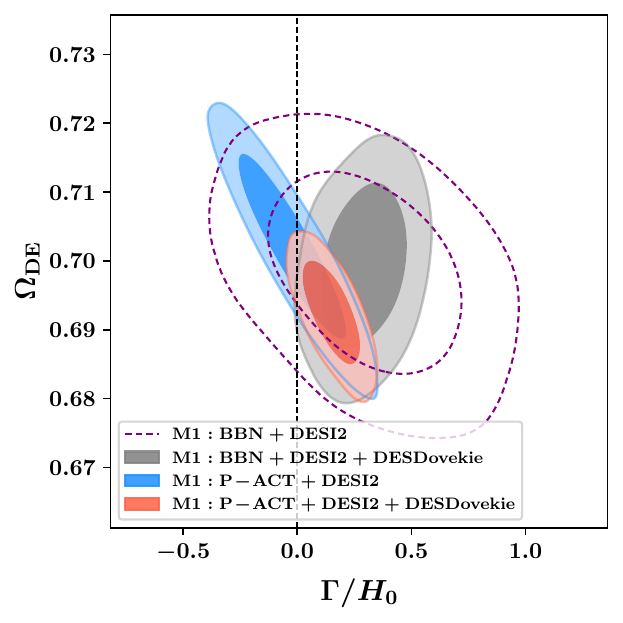}
    \includegraphics[height=0.22\textheight, width=0.325\linewidth]{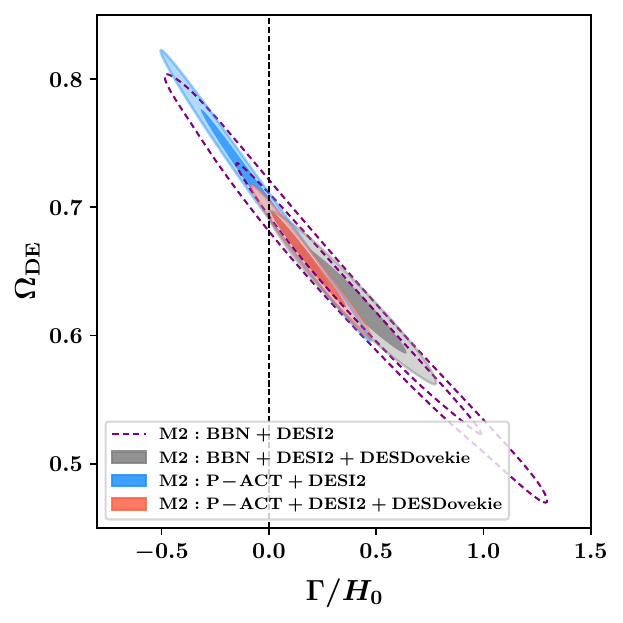}
    \includegraphics[height=0.22\textheight, width=0.325\linewidth]{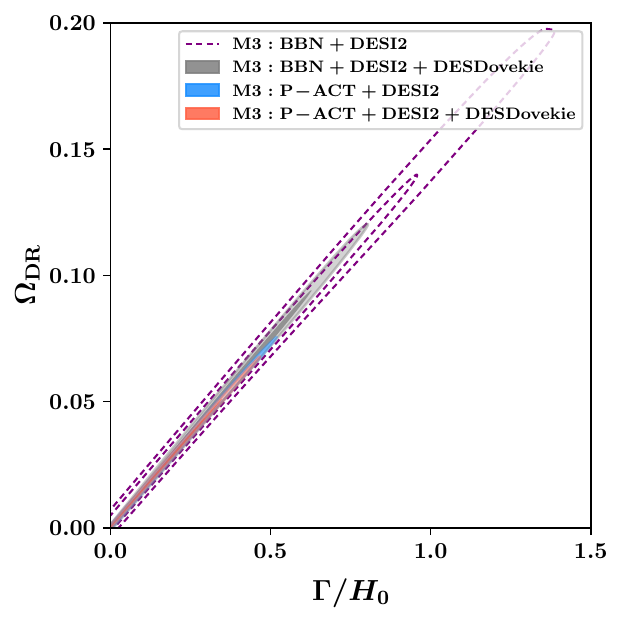}
    \caption{Marginalized Constraints on Metastable DE Models. The upper row shows the $H_0-\Omega_m$ constraints, while the lower row shows the corresponding decay-parameter planes: $\Gamma/H_0-\Omega_{\rm DE}$ for Model 1 (left panel), Model 2 (middle panel), and $\Gamma/H_0-\Omega_{\rm DR}$ for Model 3 (right panel). The vertical dashed line marks the $\Lambda$CDM limit, $\Gamma/H_0=0$.
    }
    \label{fig:mcmc}
\end{figure*}

\begin{figure*}
    \centering

    \includegraphics[height=0.22\textheight, width=0.325\linewidth]{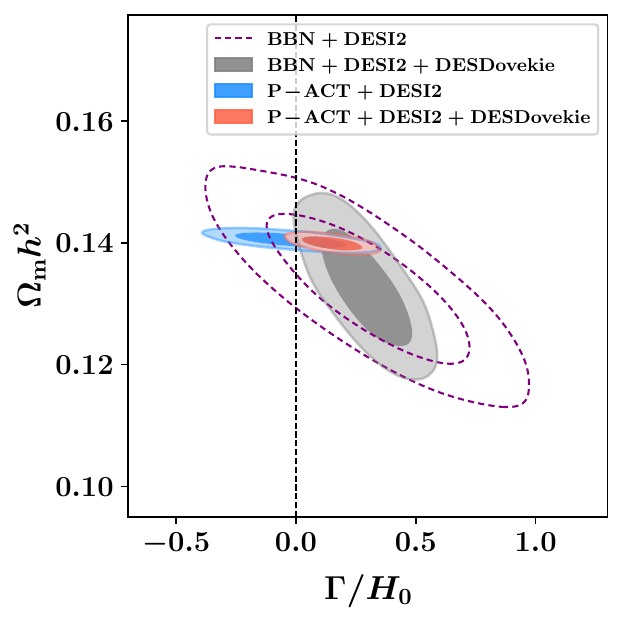}
    \begin{overpic}[height=0.22\textheight,width=0.325\linewidth]{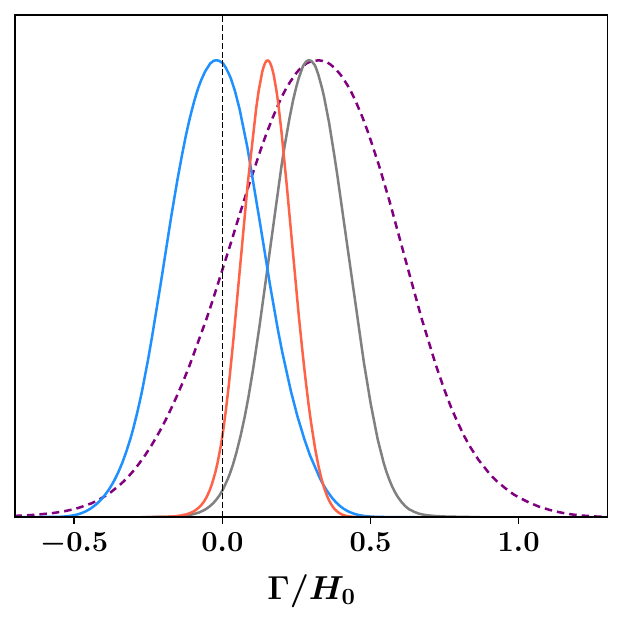}
        \put(38,88){\small\bfseries Model 1}
    \end{overpic}
    \includegraphics[height=0.22\textheight, width=0.325\linewidth]{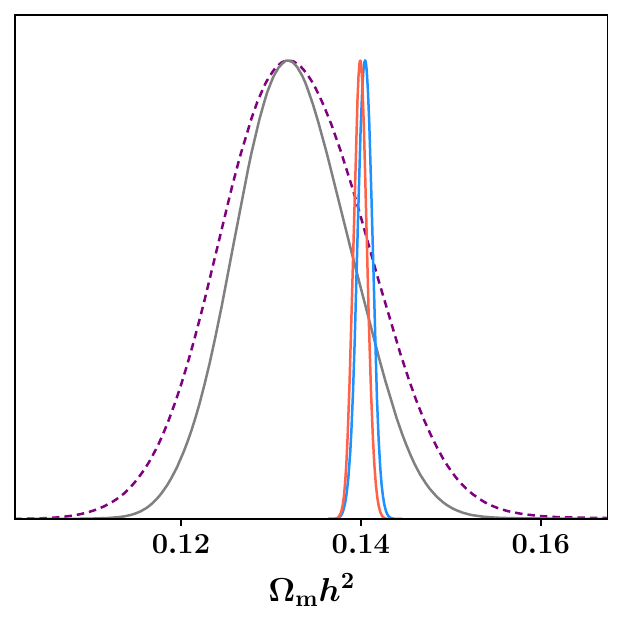}\\[0.2cm]

    \includegraphics[height=0.22\textheight, width=0.325\linewidth]{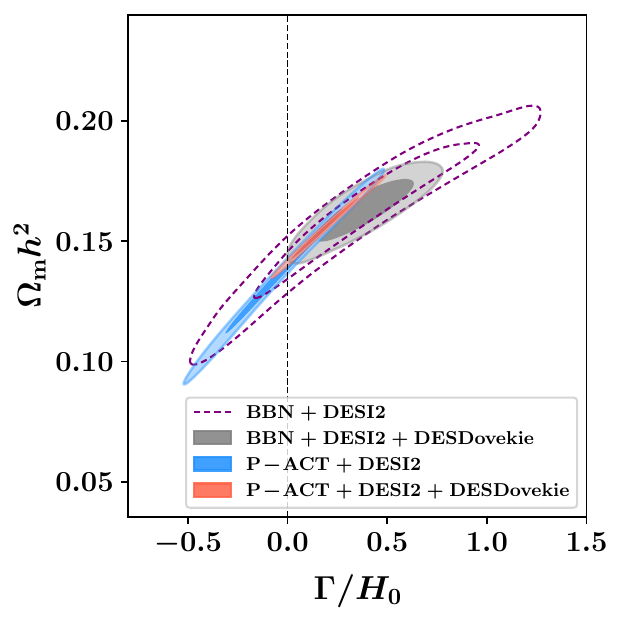}
    \begin{overpic}[height=0.22\textheight,width=0.325\linewidth]{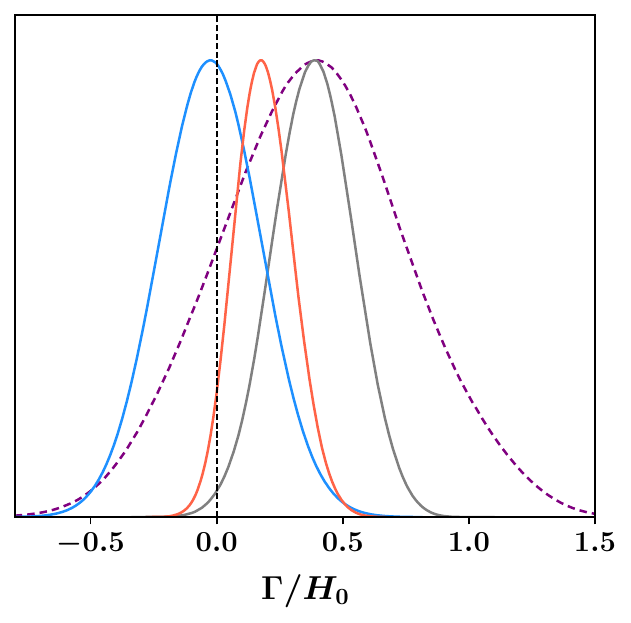}
        \put(38,88){\small\bfseries Model 2}
    \end{overpic}
    \includegraphics[height=0.22\textheight, width=0.325\linewidth]{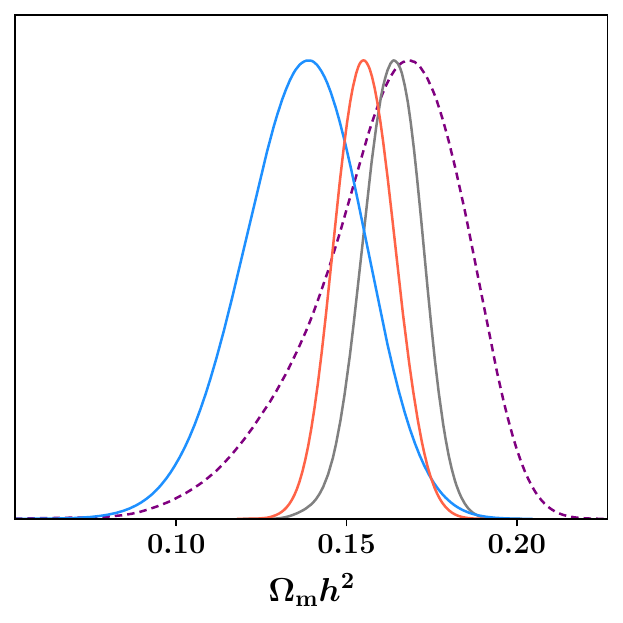}\\[0.2cm]

    \includegraphics[height=0.22\textheight, width=0.325\linewidth]{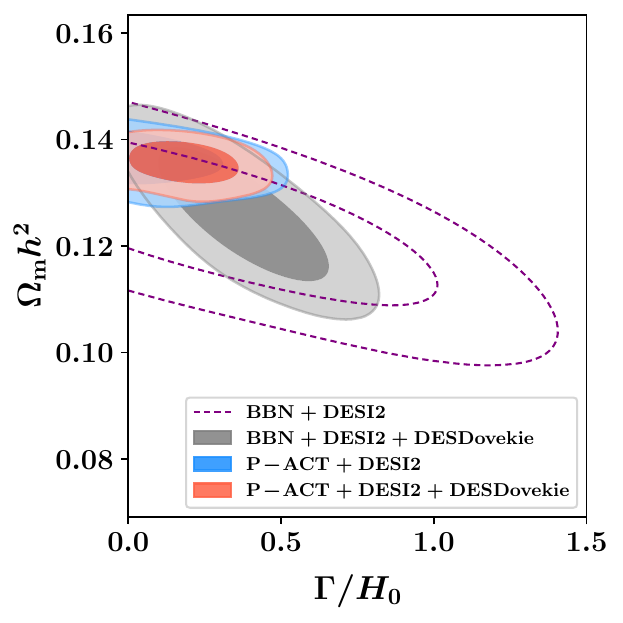}
    \begin{overpic}[height=0.22\textheight,width=0.325\linewidth]{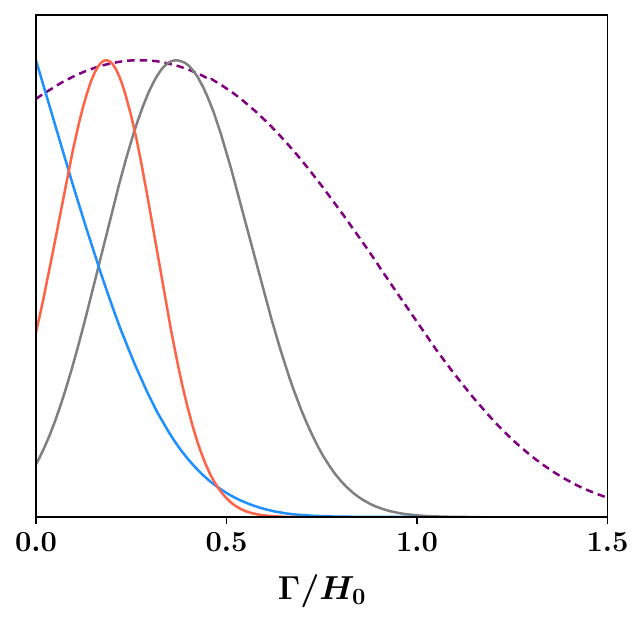}
        \put(38,88){\small\bfseries Model 3}
    \end{overpic}
    \includegraphics[height=0.22\textheight, width=0.325\linewidth]{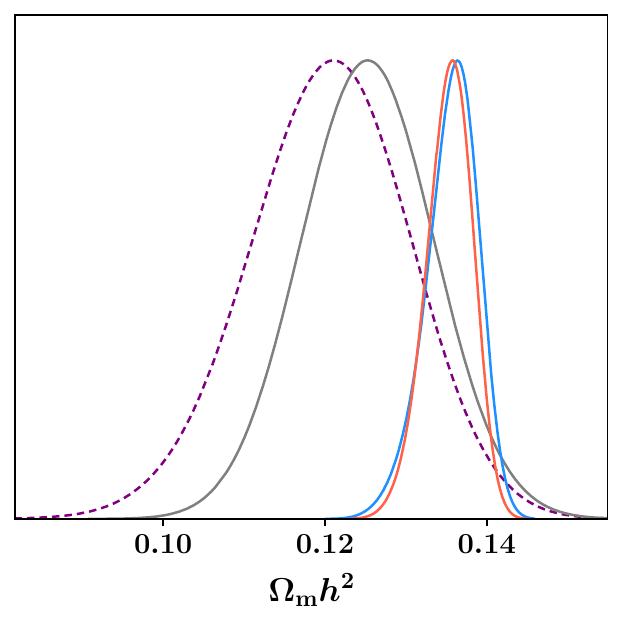}

    \caption{
    Marginalized $\Gamma/H_0-\Omega_m h^2$ constraints on Models 1--3. The center column shows the 1D posterior of $\Gamma/H_0$, and the right column shows the 1D posterior of $\Omega_m h^2$.
    }
    \label{fig:mcmc2}
\end{figure*}

Model 1 modifies the late-time background evolution through the effective DE EoS, while leaving the CMB temperature power spectrum $D_\ell^{TT}$ and the linear matter power spectrum $P(k)$ close to the $\Lambda$CDM prediction, with only a mild impact on the growth rate of structure $f\sigma_8(z)$ through the modified expansion history. Model 2 produces the largest deviations, especially in $P(k)$ and $f\sigma_8(z)$, because the decay directly changes the non-relativistic DM abundance and its perturbations. Model 3 behaves similarly to Model 1 over the shown parameter range, as the produced DR component redshifts rapidly and remains subdominant. Its distinctive signature is therefore better captured by the evolution of $\Omega_{\rm DR}$ in Fig.~\ref{fig:omega_evolution}. However, the DR component can still induce mild changes in $H(z)$ and $f\sigma_8(z)$.

For the BBN+DESI and BBN+DESI+SN-Ia combinations, we vary $\left\{\omega_b,\,  \omega_c,\, H_0,\,\Gamma/H_0 \right\},$ where $\omega_b\equiv \Omega_b h^2$ and $\omega_c\equiv \Omega_c h^2$ are the physical baryon and cold dark matter densities. For analyses involving the CMB likelihoods, we adopt the six-parameter Planck baseline framework extended by the metastable DE decay parameter and vary 
$\left\{\omega_b,\, \omega_c,\, H_0,\, 
    \tau_{\rm reio},\, n_s,\, 
    \ln(10^{10}A_s),\, \Gamma/H_0
    \right\}.$
Here $\tau_{\rm reio}$ is the optical depth to reionization, $n_s$ is the scalar spectral index, and $A_s$ is the amplitude of the primordial scalar power spectrum. The priors adopted for these parameters are summarized in Table~\ref{tab:priors}.

Parameter estimation is performed with \texttt{Cobaya}\footnote{\url{https://github.com/CobayaSampler/cobaya}} \cite{Torrado:2020dgo} using Markov Chain Monte Carlo sampling. We require the Gelman-Rubin convergence criterion to satisfy $R-1 < 0.03$.  For comparison, we also analyze the baseline $\Lambda$CDM model using the same data combinations. Best-fit points and the corresponding $\chi^2$ values are obtained with \texttt{iminuit}\footnote{\url{https://github.com/scikit-hep/iminuit.git}} \cite{James:1975dr, iminuit}, initialized at the MAP points from the MCMC chains. Model comparisons are quantified using both the relative goodness of fit and the deviance information criterion (DIC) \cite{dic, Liddle:2007fy, Grandis:2016fwl}. The relative goodness of fit with respect to the $\Lambda$CDM model is defined as $\Delta \chi^2_{\rm MAP} = \chi^2_{{\rm MAP},\,{\rm model}} - \chi^2_{{\rm MAP},\,\Lambda{\rm CDM}} \equiv -2\Delta \ln \mathcal{L}$, where the $\chi^2_{\rm MAP}$ values are evaluated at the MAP points of the corresponding models. We also compute $\Delta{\rm DIC}={\rm DIC}_{\rm model}-{\rm DIC}_{\Lambda{\rm CDM}}$ to account for the model complexity.

\squeezetable
\begin{table*}
\centering
\caption{Marginalized constraints at 68\% CL for the baseline $\Lambda$CDM and metastable DE models with BBN+DESI (+DESDovekie) data combinations.}
\label{tab:set1a_constraints}
\begin{minipage}[t]{0.85\textwidth}
\centering
\renewcommand{\arraystretch}{1.35}
\setlength{\tabcolsep}{7pt}
\resizebox{\textwidth}{!}{
\begin{tabular}{lcccc}
\toprule
\textbf{Parameter} 
& {\boldmath\textbf{$\Lambda$CDM}} 
& \textbf{Model 1} 
& \textbf{Model 2} 
& \textbf{Model 3} \\
\hline

\multicolumn{5}{c}{\textbf{BBN+DESI}} \\
\hline
{\boldmath\textbf{$H_0$}} & $68.57\pm0.60$ & $66.3^{+2.0}_{-2.2}$ & $66.6\pm2.0$ & $63.0^{+2.7}_{-2.4}$ \\
{\boldmath\textbf{$\Omega_b h^2$}} & $0.02222^{+0.00051}_{-0.00057}$ & $0.02221\pm0.00055$ & $0.02218\pm0.00054$ & $0.02225\pm0.00054$ \\
{\boldmath\textbf{$\Omega_c h^2$}} & $0.1169\pm0.0048$ & $0.1099\pm0.0081$ & $0.139^{+0.026}_{-0.016}$ & $0.1075\pm0.0077$ \\
{\boldmath\textbf{$\Gamma/H_0$}} & -- & $0.31\pm0.28$ & $0.38\pm0.37$ & $<0.701$ \\
{\boldmath\textbf{$\Omega_m$}} & $0.2972\pm0.0083$ & $0.3015\pm0.0096$ & $0.369^{+0.075}_{-0.063}$ & $0.3032\pm0.0095$ \\
{\boldmath\textbf{$\Omega_{\rm DR}$}} & -- & -- & -- & $<0.103$ \\
\hline
\multicolumn{5}{c}{\textbf{BBN+DESI+DES-Dovekie}} \\
\hline
{\boldmath\textbf{$H_0$}} & $68.63\pm0.60$ & $66.4\pm1.1$ & $66.4\pm1.1$ & $64.5\pm1.5$ \\
{\boldmath\textbf{$\Omega_b h^2$}} & $0.02217\pm0.00056$ & $0.02219\pm0.00055$ & $0.02218\pm0.00057$ & $0.02219\pm0.00056$ \\
{\boldmath\textbf{$\Omega_c h^2$}} & $0.1214\pm0.0044$ & $0.1098^{+0.0059}_{-0.0066}$ & $0.1404^{+0.0088}_{-0.0080}$ & $0.1097\pm0.0065$ \\
{\boldmath\textbf{$\Gamma/H_0$}} & -- & $0.29\pm0.12$ & $0.38\pm0.16$ & $0.39^{+0.16}_{-0.20}$ \\
{\boldmath\textbf{$\Omega_m$}} & $0.3062\pm0.0075$ & $0.3007\pm0.0078$ & $0.371\pm0.027$ & $0.3011\pm0.0085$ \\
{\boldmath\textbf{$\Omega_{\rm DR}$}} & -- & -- & -- & $0.058^{+0.024}_{-0.030}$ \\
\bottomrule
\end{tabular}
}
\end{minipage}
\end{table*}


\squeezetable
\begin{table*}
\centering
\caption{Marginalized constraints at 68\% CL for the baseline $\Lambda$CDM and metastable DE models with Planck+DESI (+DESDovekie) data combinations.}
\label{tab:set2a_constraints}
\begin{minipage}[t]{0.85\textwidth}
\centering
\renewcommand{\arraystretch}{1.35}
\setlength{\tabcolsep}{7pt}
\resizebox{\textwidth}{!}{
\begin{tabular}{lcccc}
\toprule
\textbf{Parameter} 
& {\boldmath\textbf{$\Lambda$CDM}} 
& \textbf{Model 1} 
& \textbf{Model 2} 
& \textbf{Model 3} \\
\hline
\multicolumn{5}{c}{\textbf{Planck+DESI}} \\
\hline
{\boldmath\textbf{$\log(10^{10}A_s)$}} & $3.048^{+0.013}_{-0.015}$ & $3.046\pm0.014$ & $3.046^{+0.013}_{-0.015}$ & $3.049\pm0.014$ \\
{\boldmath\textbf{$n_s$}} & $0.9683\pm0.0033$ & $0.9675\pm0.0035$ & $0.9677^{+0.0036}_{-0.0032}$ & $0.9687\pm0.0036$ \\
{\boldmath\textbf{$H_0$}} & $68.20\pm0.29$ & $69.0\pm1.1$ & $68.9\pm1.1$ & $66.7^{+1.1}_{-0.80}$ \\
{\boldmath\textbf{$\Omega_b h^2$}} & $0.02233\pm0.00012$ & $0.02230\pm0.00013$ & $0.02231\pm0.00012$ & $0.02234\pm0.00012$ \\
{\boldmath\textbf{$\Omega_c h^2$}} & $0.11755\pm0.00063$ & $0.11787\pm0.00079$ & $0.105\pm0.018$ & $0.11730\pm0.00066$ \\
{\boldmath\textbf{$\tau_{\rm reio}$}} & $0.0594\pm0.0072$ & $0.0582^{+0.0066}_{-0.0077}$ & $0.0582^{+0.0065}_{-0.0074}$ & $0.0603\pm0.0072$ \\
{\boldmath\textbf{$\Gamma/H_0$}} & -- & $-0.11\pm0.15$ & $-0.13^{+0.17}_{-0.20}$ & $<0.146$ \\
{\boldmath\textbf{$\Omega_m$}} & $0.3021\pm0.0037$ & $0.2964\pm0.0088$ & $0.272\pm0.046$ & $0.3077^{+0.0051}_{-0.0066}$ \\
{\boldmath\textbf{$\Omega_{\rm DR}$}} & -- & -- & -- & $<0.0216$ \\
\hline 
\multicolumn{5}{c}{\textbf{Planck+DESI+DES-Dovekie}} \\
\hline
{\boldmath\textbf{$\log(10^{10}A_s)$}} & $3.047\pm0.014$ & $3.050\pm0.014$ & $3.049\pm0.014$ & $3.050\pm0.014$ \\
{\boldmath\textbf{$n_s$}} & $0.9676\pm0.0033$ & $0.9691\pm0.0034$ & $0.9686\pm0.0035$ & $0.9689\pm0.0034$ \\
{\boldmath\textbf{$H_0$}} & $68.08\pm0.28$ & $67.43\pm0.55$ & $67.47\pm0.53$ & $66.45^{+0.86}_{-0.72}$ \\
{\boldmath\textbf{$\Omega_b h^2$}} & $0.02230\pm0.00012$ & $0.02235^{+0.00013}_{-0.00011}$ & $0.02233\pm0.00013$ & $0.02235\pm0.00012$ \\
{\boldmath\textbf{$\Omega_c h^2$}} & $0.11782\pm0.00062$ & $0.11728\pm0.00071$ & $0.1285\pm0.0081$ & $0.11727\pm0.00070$ \\
{\boldmath\textbf{$\tau_{\rm reio}$}} & $0.0586\pm0.0069$ & $0.0606^{+0.0064}_{-0.0076}$ & $0.0604^{+0.0068}_{-0.0076}$ & $0.0606\pm0.0072$ \\
{\boldmath\textbf{$\Gamma/H_0$}} & -- & $0.111\pm0.079$ & $0.14\pm0.11$ & $0.158^{+0.061}_{-0.13}$ \\
{\boldmath\textbf{$\Omega_m$}} & $0.3037\pm0.0036$ & $0.3086\pm0.0051$ & $0.333\pm0.023$ & $0.3095\pm0.0047$ \\
{\boldmath\textbf{$\Omega_{\rm DR}$}} & -- & -- & -- & $0.0233^{+0.0094}_{-0.019}$ \\
\bottomrule
\end{tabular}
}
\end{minipage}
\end{table*}


\squeezetable
\begin{table*}
\centering
\caption{Marginalized constraints at 68\% CL for the baseline $\Lambda$CDM and metastable DE models with P-ACT+DESI (+DESDovekie) data combinations.}
\label{tab:set3a_constraints}
\begin{minipage}[t]{0.85\textwidth}
\centering
\renewcommand{\arraystretch}{1.35}
\setlength{\tabcolsep}{7pt}
\resizebox{\textwidth}{!}{
\begin{tabular}{lcccc}
\toprule
\textbf{Parameter} 
& {\boldmath\textbf{$\Lambda$CDM}} 
& \textbf{Model 1} 
& \textbf{Model 2} 
& \textbf{Model 3} \\
\hline
\multicolumn{5}{c}{\textbf{P-ACT+DESI}} \\
\hline
{\boldmath\textbf{$\log(10^{10}A_s)$}} & $3.063\pm0.011$ & $3.063\pm0.011$ & $3.063\pm0.011$ & $3.065\pm0.011$ \\
{\boldmath\textbf{$n_s$}} & $0.9752\pm0.0030$ & $0.9751\pm0.0032$ & $0.9752\pm0.0031$ & $0.9760\pm0.0031$ \\
{\boldmath\textbf{$H_0$}} & $68.49\pm0.28$ & $68.7\pm1.1$ & $68.7\pm1.0$ & $66.6^{+1.2}_{-1.0}$ \\
{\boldmath\textbf{$\Omega_b h^2$}} & $0.02255\pm0.00010$ & $0.02256\pm0.00011$ & $0.02256\pm0.00010$ & $0.02257\pm0.00010$ \\
{\boldmath\textbf{$\Omega_c h^2$}} & $0.11721\pm0.00068$ & $0.11724\pm0.00085$ & $0.114\pm0.017$ & $0.11674\pm0.00075$ \\
{\boldmath\textbf{$\tau_{\rm reio}$}} & $0.0612^{+0.0058}_{-0.0066}$ & $0.0613^{+0.0056}_{-0.0066}$ & $0.0612\pm0.0064$ & $0.0617\pm0.0064$ \\
{\boldmath\textbf{$\Gamma/H_0$}} & -- & $-0.02\pm0.15$ & $-0.02\pm0.19$ & $<0.198$ \\
{\boldmath\textbf{$\Omega_m$}} & $0.2994\pm0.0037$ & $0.2981\pm0.0087$ & $0.292\pm0.044$ & $0.3059^{+0.0061}_{-0.0075}$ \\
{\boldmath\textbf{$\Omega_{\rm DR}$}} & -- & -- & -- & $<0.0294$  \\
\hline
\multicolumn{5}{c}{\textbf{P-ACT+DESI+DES-Dovekie}} \\
\hline
{\boldmath\textbf{$\log(10^{10}A_s)$}} & $3.0623^{+0.0098}_{-0.011}$ & $3.066\pm0.011$ & $3.064\pm0.011$ & $3.066\pm 0.011 $ \\
{\boldmath\textbf{$n_s$}} & $0.9745\pm0.0031$ & $0.9764\pm0.0031$ & $0.9756\pm0.0031$ & $0.9763\pm 0.0032 $ \\
{\boldmath\textbf{$H_0$}} & $68.37\pm0.27$ & $67.47\pm0.55$ & $67.53\pm0.55$ & $66.30^{+0.97}_{-0.86}$ \\
{\boldmath\textbf{$\Omega_b h^2$}} & $0.02254\pm0.00010$ & $0.02258\pm0.00011$ & $0.02257\pm0.00011$ & $0.02258\pm 0.00011$ \\
{\boldmath\textbf{$\Omega_c h^2$}} & $0.11750\pm0.00065$ & $0.11671\pm0.00077$ & $0.1321\pm0.0082$ & $0.11670\pm 0.00077$ \\
{\boldmath\textbf{$\tau_{\rm reio}$}} & $0.0612^{+0.0055}_{-0.0065}$ & $0.0619^{+0.0058}_{-0.0068}$ & $0.0615\pm0.0063$ & $0.0621\pm 0.0064$ \\
{\boldmath\textbf{$\Gamma/H_0$}} & -- & $0.151\pm0.082$ & $0.19\pm0.11$ & $0.204^{+0.093}_{-0.14}$ \\
{\boldmath\textbf{$\Omega_m$}} & $0.3010\pm0.0036$ & $0.3074\pm0.0050$ & $0.341\pm0.023$ & $0.3081\pm 0.0051$ \\
{\boldmath\textbf{$\Omega_{\rm DR}$}} & -- & -- & -- & $0.030^{+0.014}_{-0.019}$ \\
\bottomrule
\end{tabular}
}
\end{minipage}
\end{table*}

\section{Results and Discussions \label{sec:results}}

We present the parameter constraints for the three metastable DE scenarios in Fig.~\ref{fig:mcmc}. The upper panels show the $\Omega_m-H_0$ constraints, while the lower panels show the corresponding decay-parameter planes: $\Gamma/H_0-\Omega_{\rm DE}$ for Models 1 and 2, and $\Gamma/H_0-\Omega_{\rm DR}$ for Model 3. The marginalized constraints on $\Gamma/H_0$ and $\Omega_m h^2$ are shown in Fig.~\ref{fig:mcmc2}. The numerical constraints are summarized in Tables~\ref{tab:set1a_constraints}, \ref{tab:set2a_constraints}, and \ref{tab:set3a_constraints}, where the baseline $\Lambda$CDM results are included for comparison. In the main text, we focus on the DES-Dovekie SN-Ia combination, while the corresponding Union3 and PantheonPlus results are given in Appendix~\ref{app:snia}.

\begin{enumerate}[left=0pt]
    \item The BBN+DESI constraints are generally broad in the $\Omega_m-H_0$ plane. Adding SN-Ia data tightens the contours and tends to shift the posterior toward $\Gamma/H_0>0$, indicating that late-time distance data can accommodate a decaying DE density. By contrast, CMB-based combinations give tighter constraints and pull the decay parameter closer to the $\Lambda$CDM limit, $\Gamma/H_0=0$, because of the precise early-universe constraints on the matter density and radiation sector.
    \item For Model 1, where the DE density decays exponentially, $\Gamma/H_0$ modifies the late-time background expansion through the effective DE EoS. $\Gamma/H_0>0$ values correspond to faster DE decay, implying a larger DE density in the past for fixed $\Omega_{\rm DE}$ today. This produces the correlation seen in the $\Gamma/H_0-\Omega_{\rm DE}$ plane. Although the BBN+DESI+SN-Ia combinations mildly favour positive decay rates, the $\Gamma/H_0=0$ limit remains within the $3\sigma$ CL. However, the CMB-informed constraints shift the posterior closer to zero, with $\Gamma/H_0=0$ lying within the $2\sigma$ region.
    \item For Model 2, where DE decays into non-baryonic DM, the decay affects both the background matter abundance and the evolution of DM perturbations. As a result, the $\Omega_m-H_0$ contours are broader and shifted toward larger $\Omega_m$, especially for the BBN+DESI-based combinations. Positive $\Gamma/H_0$ transfers energy from DE to DM, increasing the effective matter abundance and shifting $\Omega_m h^2$ relative to $\Lambda$CDM. The $\Gamma/H_0-\Omega_{\rm DE}$ plane shows a strong anti-correlation, since a larger decay rate can be compensated by a lower present-day $\Omega_{\rm DE}$ to preserve the expansion history. 
    \item For Model 3, where DE decays into DR, the produced DR component has no primordial abundance and is sourced only by DE decay, so the model is restricted to $\Gamma>0$. The BBN+DESI-based combinations allow positive $\Gamma/H_0$ values and hence a nonzero $\Omega_{\rm DR}$. Adding SN-Ia data tightens the contours while still permitting a nonzero DR abundance at $2\sigma$. The $\Gamma/H_0-\Omega_{\rm DR}$ plane shows a positive correlation, since larger decay rates produce larger DR abundances. Once CMB information is included, the allowed region becomes significantly narrower, with both $\Gamma/H_0$ and $\Omega_{\rm DR}$ restricted to smaller values. This reflects the strong CMB sensitivity to additional relativistic components through their impact on the early-time radiation sector and acoustic scale.
\end{enumerate}
Overall, the marginalized constraints show that positive values of $\Gamma/H_0$ are allowed and mildly preferred in some late-time data combinations, although this trend remains dataset-dependent. The three decay channels lead to distinct signatures: background-level changes in Model 1, matter-sector shifts in Model 2, and a DR contribution in Model 3, with the latter strongly constrained once CMB data are included.

\begin{figure*}
    \centering
    \includegraphics[width=0.325\linewidth]{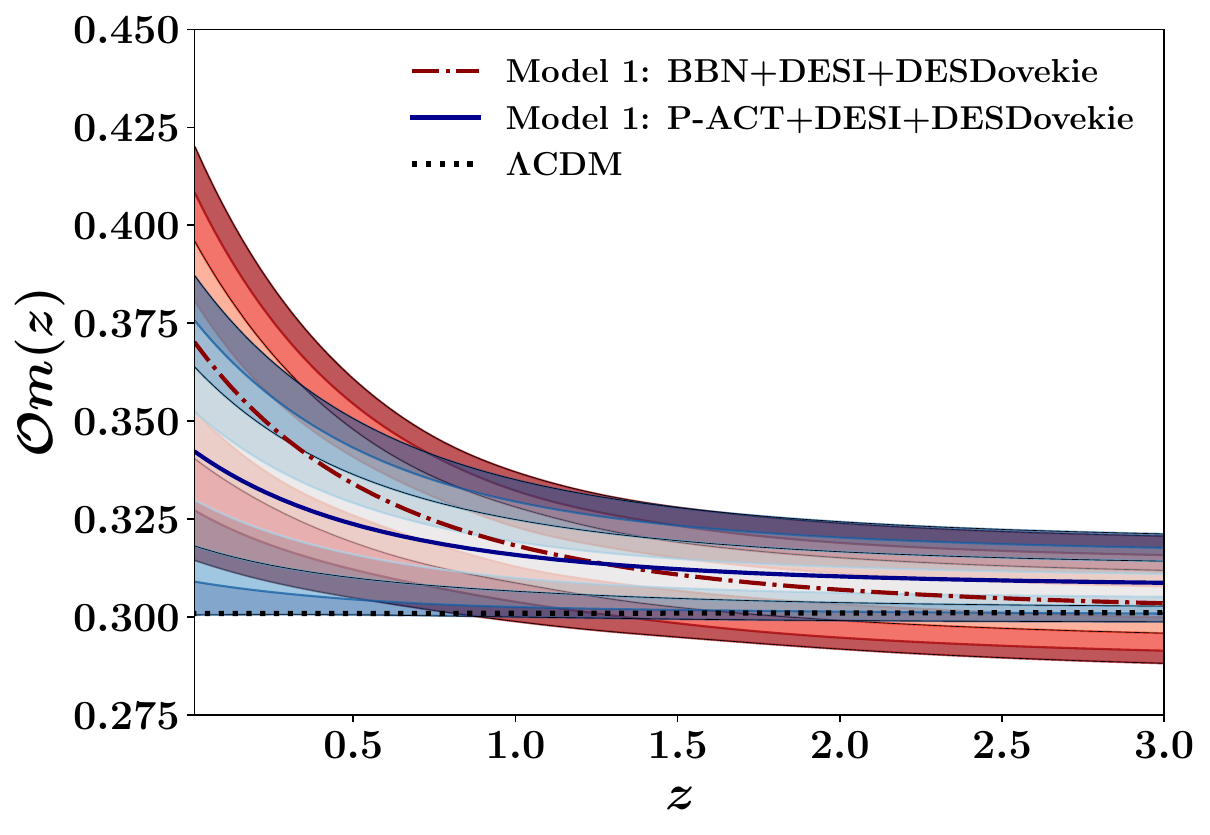}
    \includegraphics[width=0.325\linewidth]{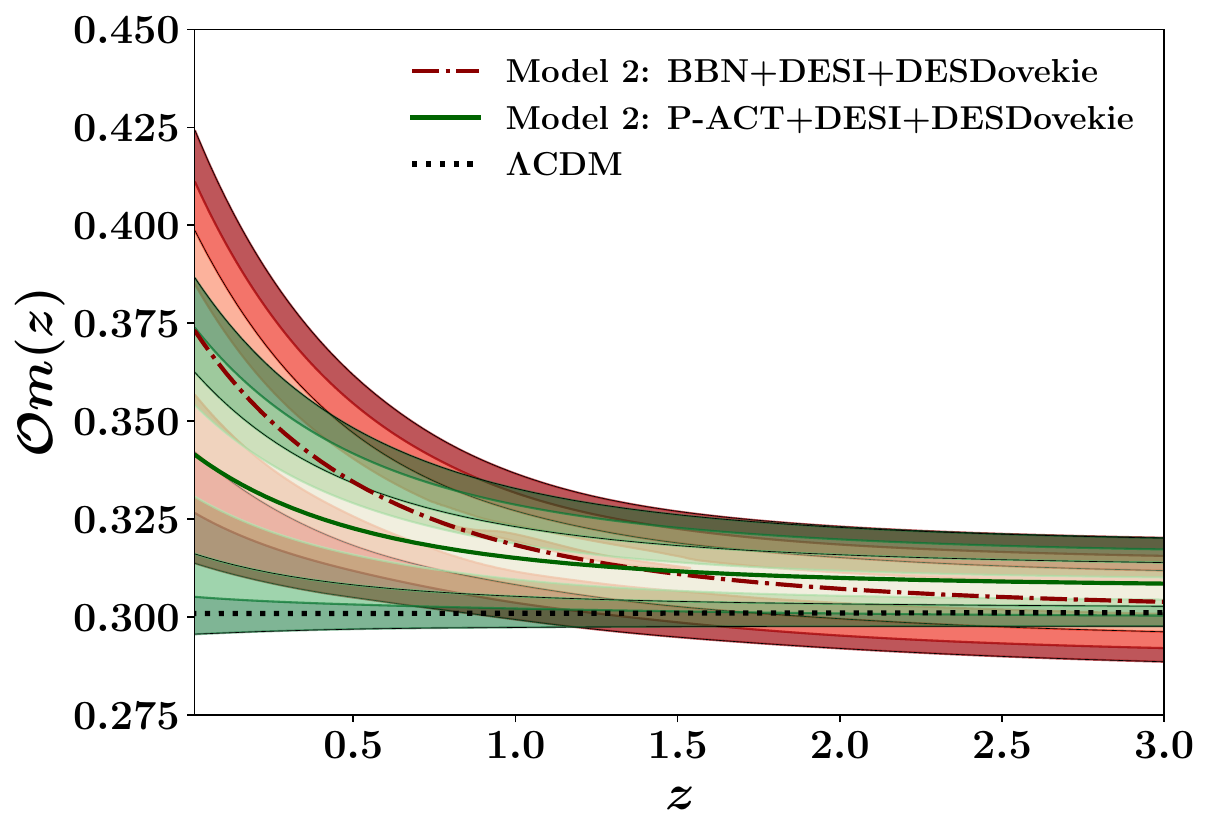}
    \includegraphics[width=0.325\linewidth]{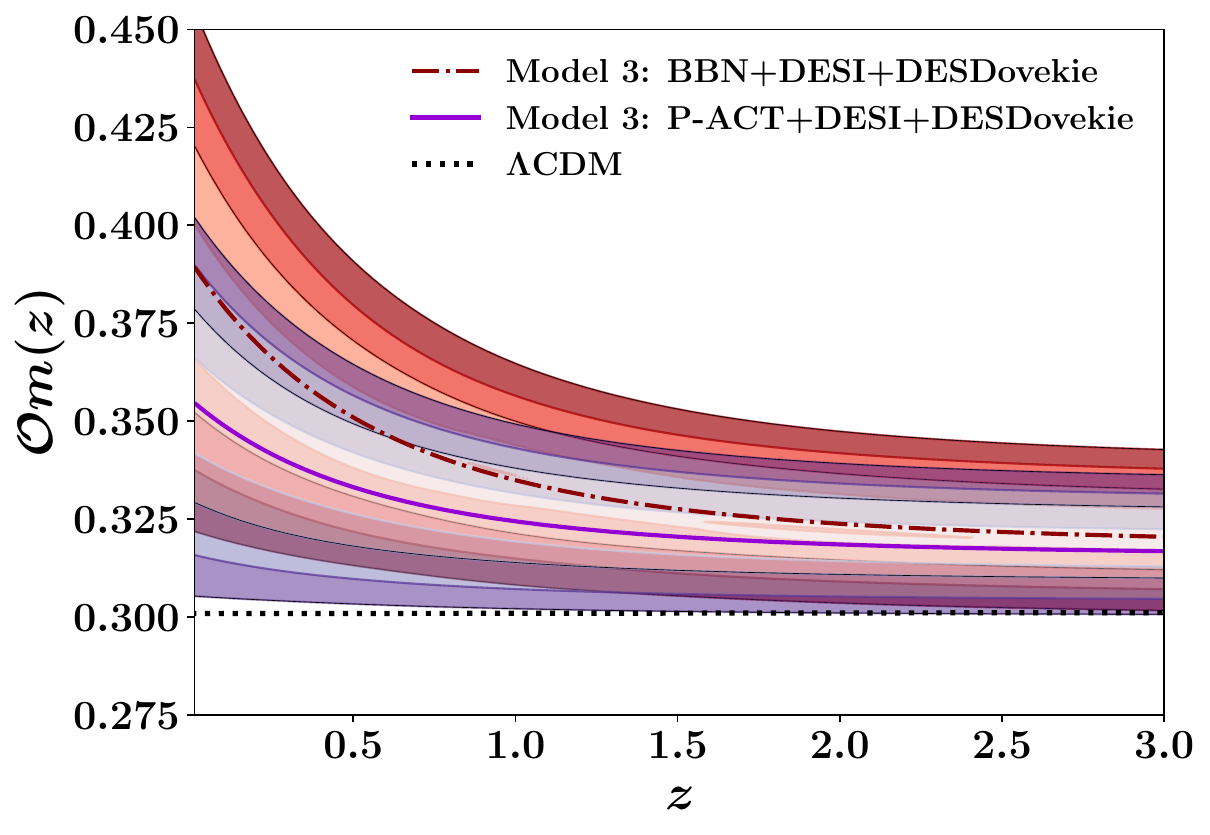}
    \caption{Reconstructed $Om(z)$ diagnostic for the metastable DE models using representative BBN+DESI+DES-Dovekie and P-ACT+DESI+DES-Dovekie combinations. The best-fit $\Lambda$CDM prediction is shown for comparison.}
    \label{fig:omz}
\end{figure*}

\begin{figure*}
    \centering
    \includegraphics[width=0.325\linewidth]{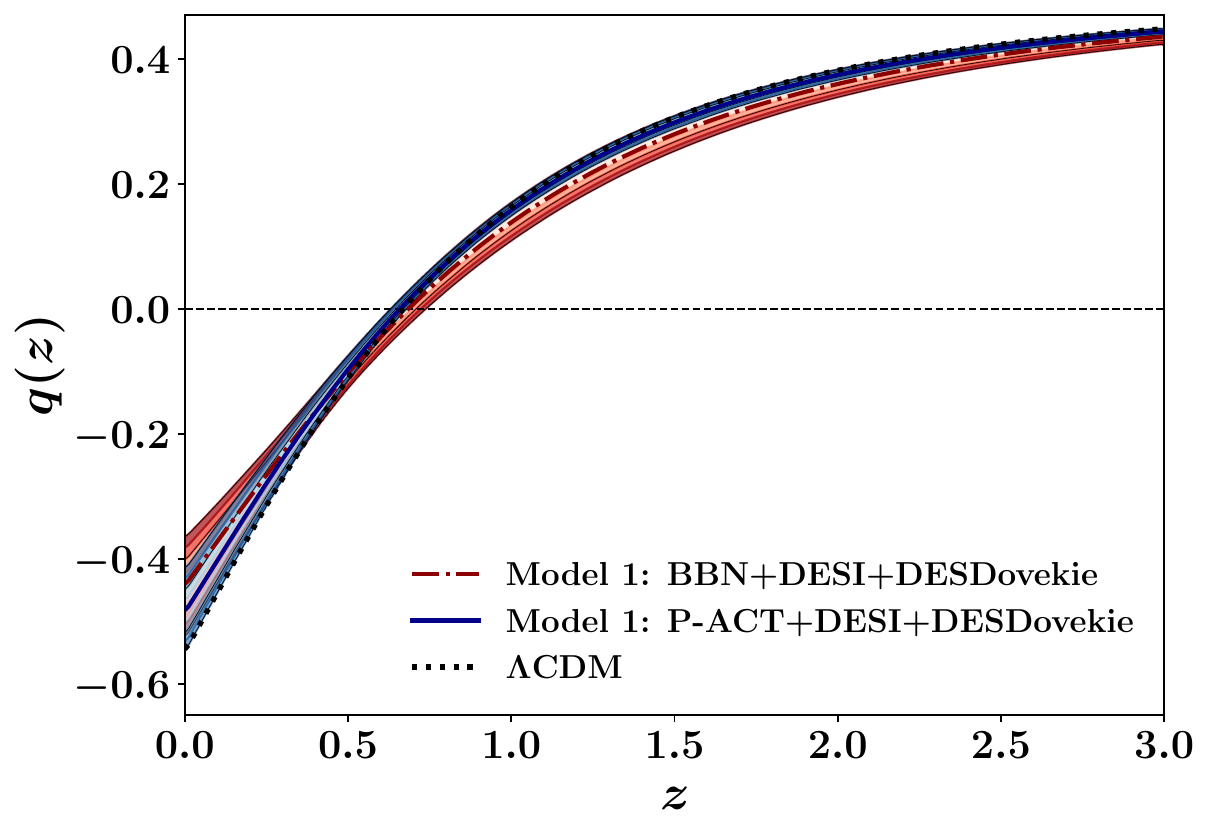}
    \includegraphics[width=0.325\linewidth]{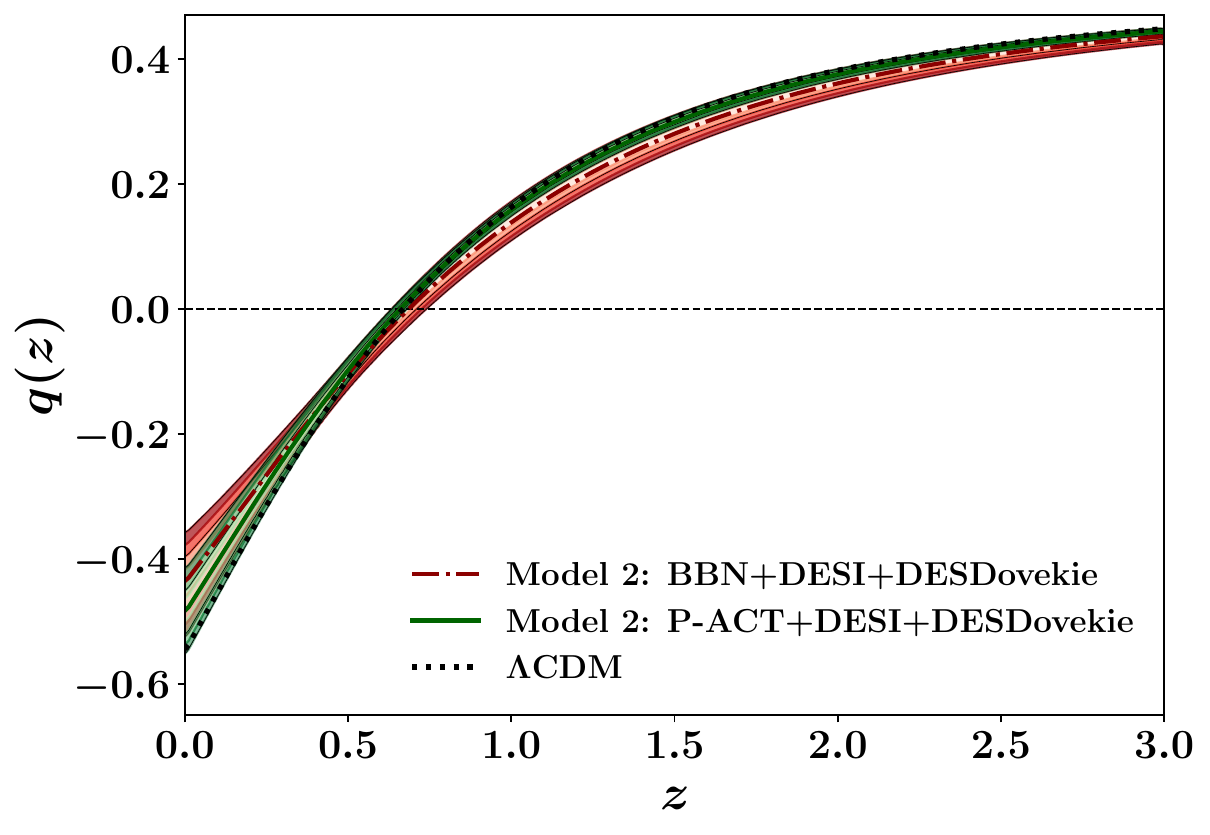}
    \includegraphics[width=0.325\linewidth]{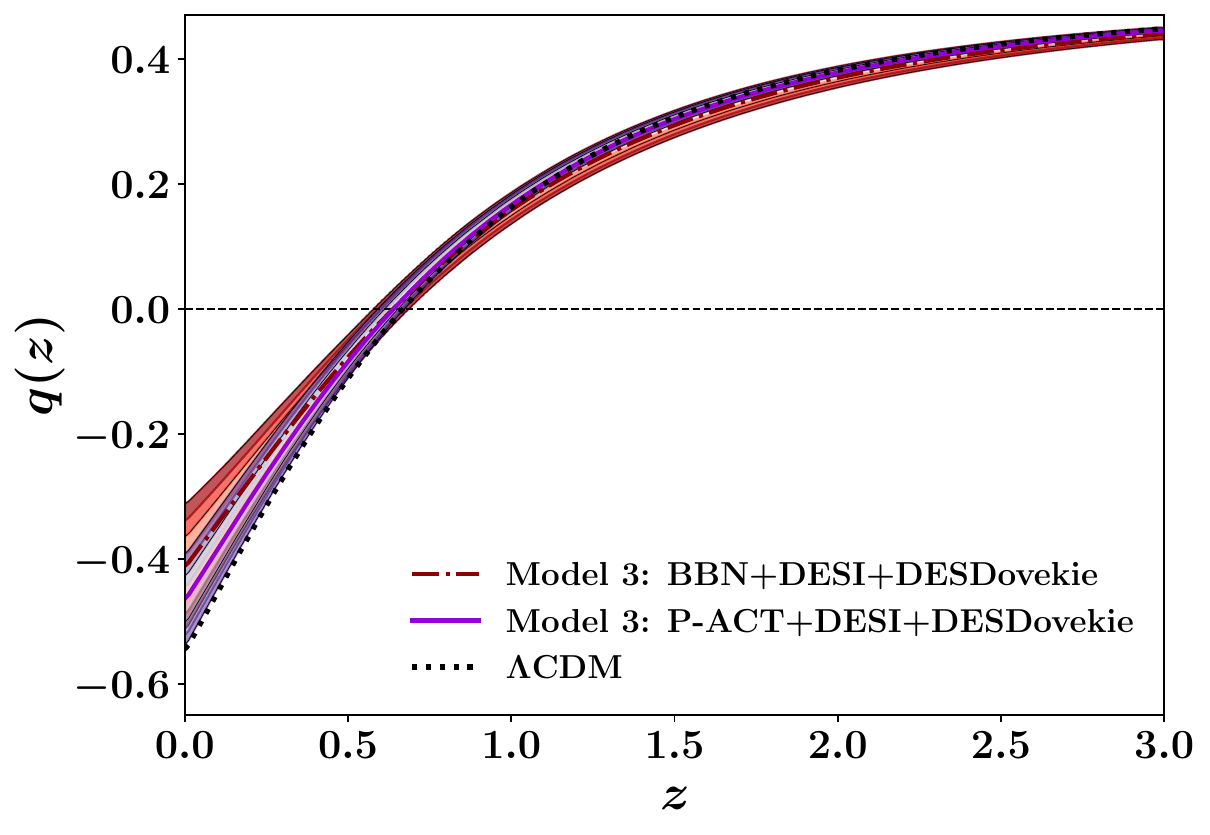}
    \caption{Reconstructed evolution of the deceleration parameter $q(z)$ for the metastable DE models.}
    \label{fig:qz}
\end{figure*}

\begin{figure*}
    \centering
    \includegraphics[width=0.325\linewidth]{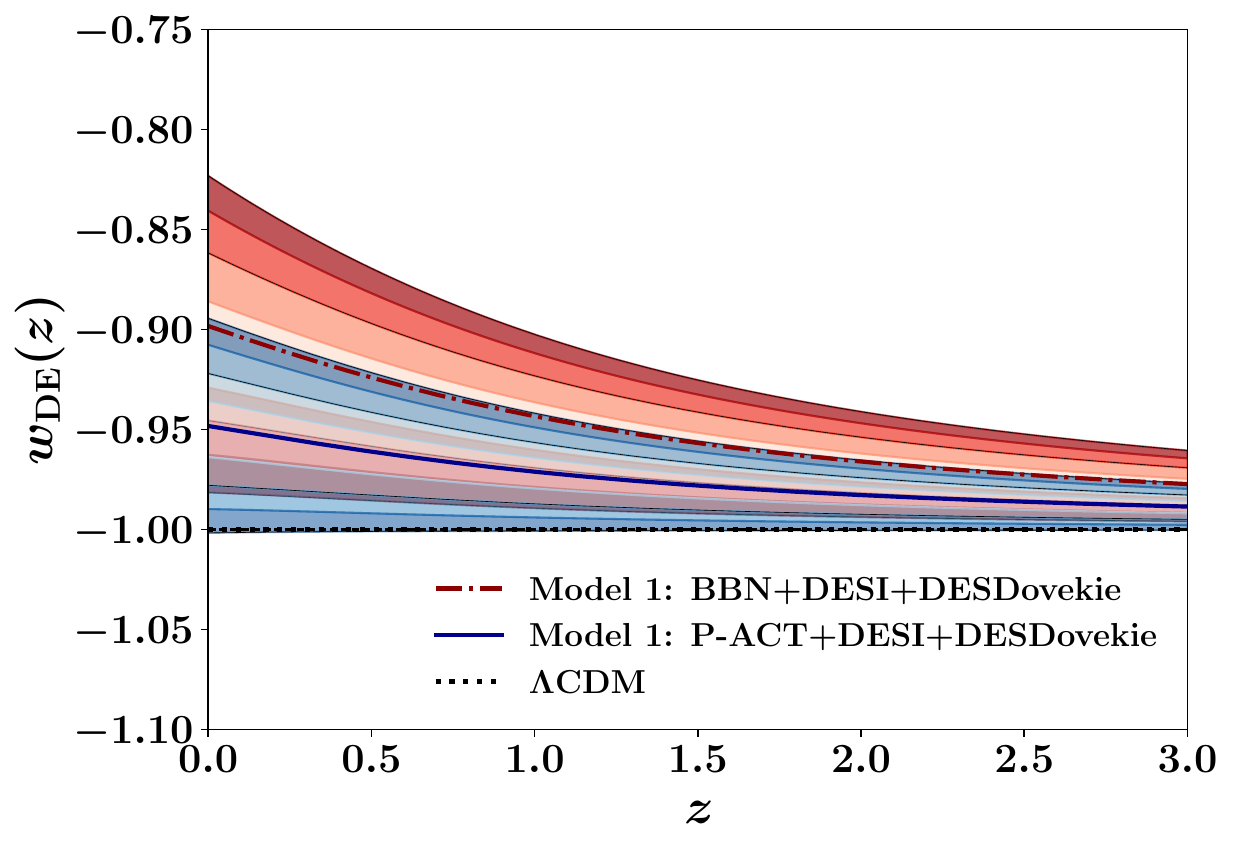}
    \includegraphics[width=0.325\linewidth]{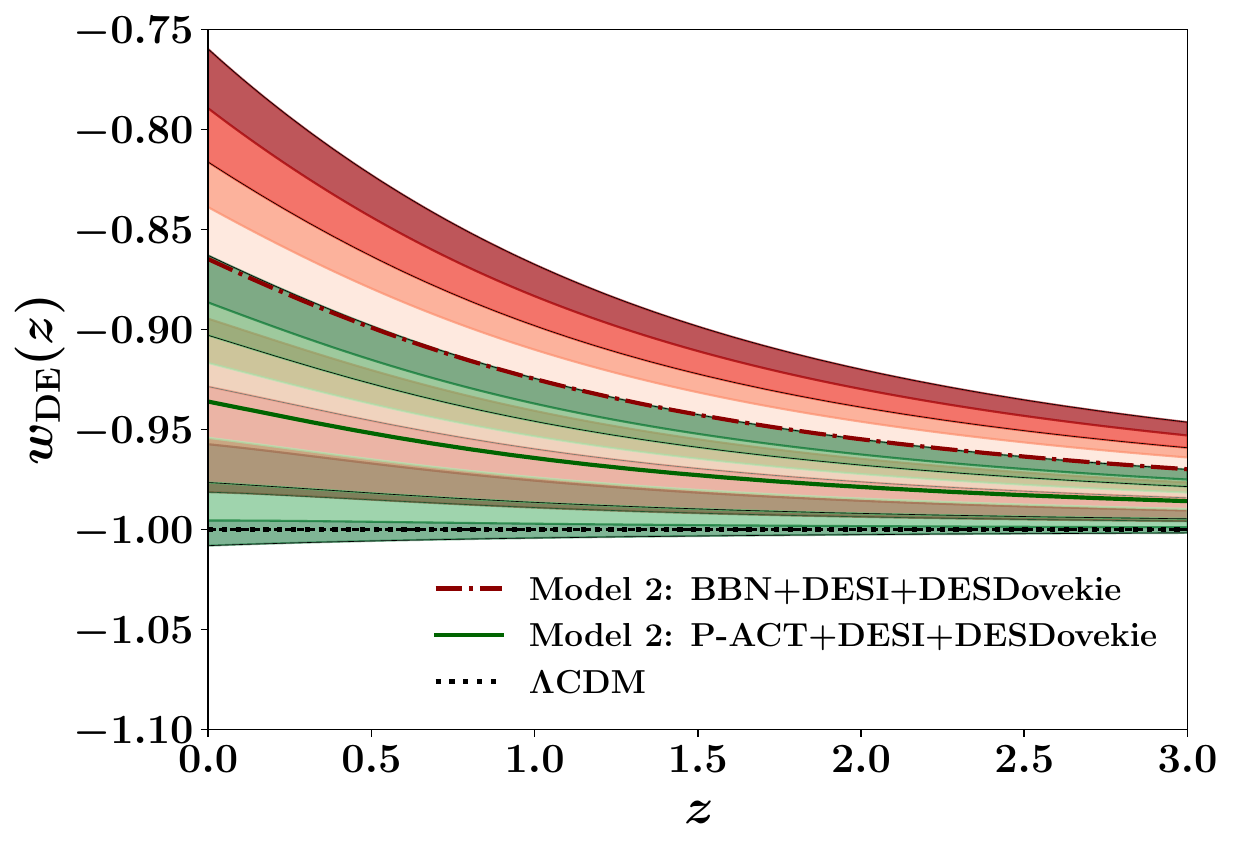}
    \includegraphics[width=0.325\linewidth]{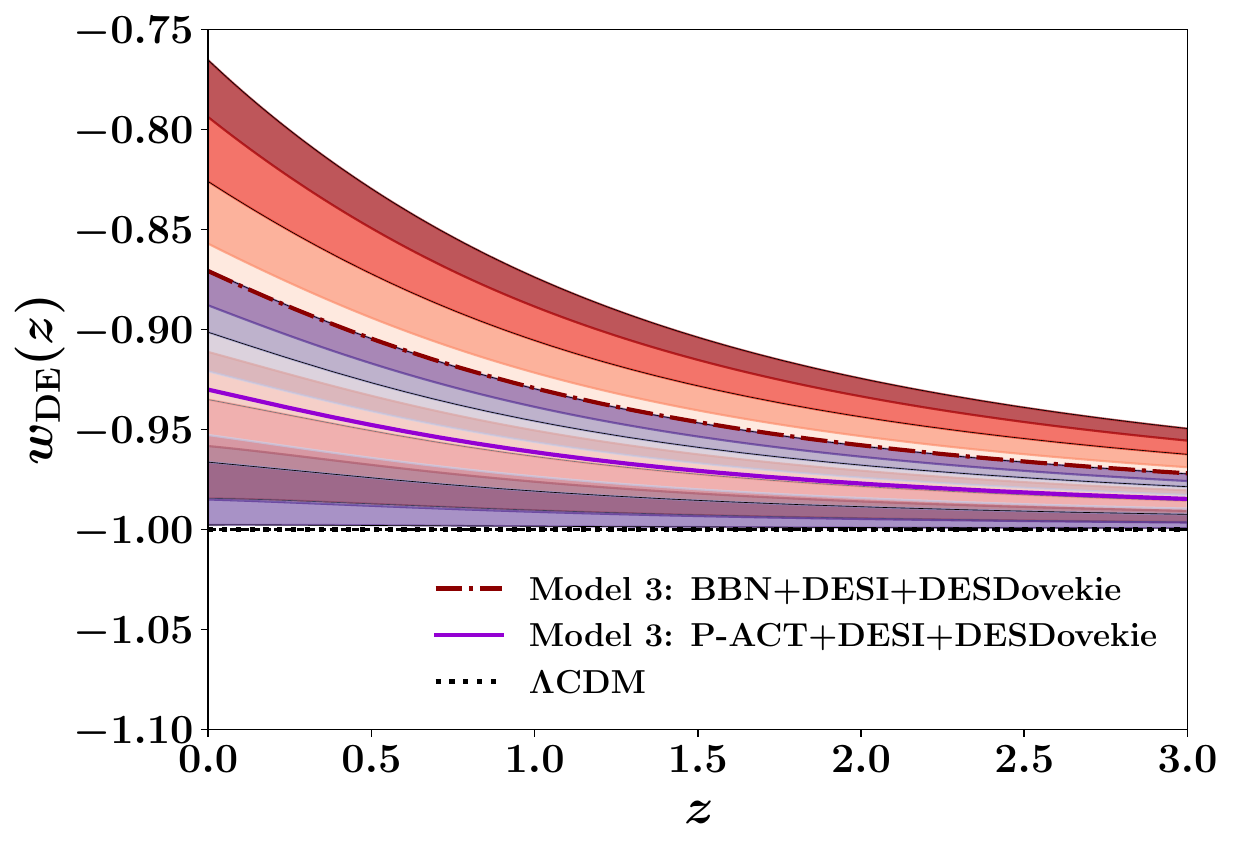}
    \caption{Reconstructed effective DE equation of state $w_{\rm DE}(z)$ for the three metastable DE models. The $\Lambda$CDM model, $w_{\rm DE}=-1$, is shown for comparison.}
    \label{fig:wdez}
\end{figure*}

\section{Reconstructed Background and Growth Observables}
\label{sec:evo}

We reconstruct several derived background and growth observables from the MCMC posterior samples to examine how the metastable DE scenarios modify the late-time expansion and structure-growth histories. We first consider the $\mathcal{O}m(z)$ diagnostic \cite{Sahni:2008xx}, 
\begin{equation} 
\mathcal{O}m(z) = \frac{{H^2(z)}/{H_0^2}-1}{(1+z)^3-1} \, .
\end{equation} 
For a spatially flat $\Lambda$CDM model, $\mathcal{O}m(z)$ is constant and equal to $\Omega_{m0}$. Therefore, any redshift dependence in $\mathcal{O}m(z)$ indicates a departure from the $\Lambda$CDM scenario. Figure~\ref{fig:omz} shows the reconstructed $\mathcal{O}m(z)$ diagnostic. 

For the BBN+DESI+DES-Dovekie combination, the reconstructed $\mathcal{O}m(z)$ bands show a clear low-redshift departure from the constant $\Lambda$CDM prediction. Over part of the low-$z$ range, the best-fit $\Lambda$CDM curve lies outside the $\sim 2\sigma$ region for all three metastable DE models. This suggests that late-time BAO and SN-Ia distance data favour an evolving DE sector within these models. When CMB information is included, the reconstructed bands become significantly narrower, reflecting the stronger constraints on $\Omega_m h^2$ and the expansion history. The remaining deviation is model dependent: Model 1 lies close to the $2\sigma$ boundary, Model 2 remains broadly consistent with $\Lambda$CDM within the $2\sigma$ region, while Model 3 still shows a noticeable residual departure from the $\Lambda$CDM best-fit.

\begin{figure*}
    \centering
    \includegraphics[width=0.325\linewidth]{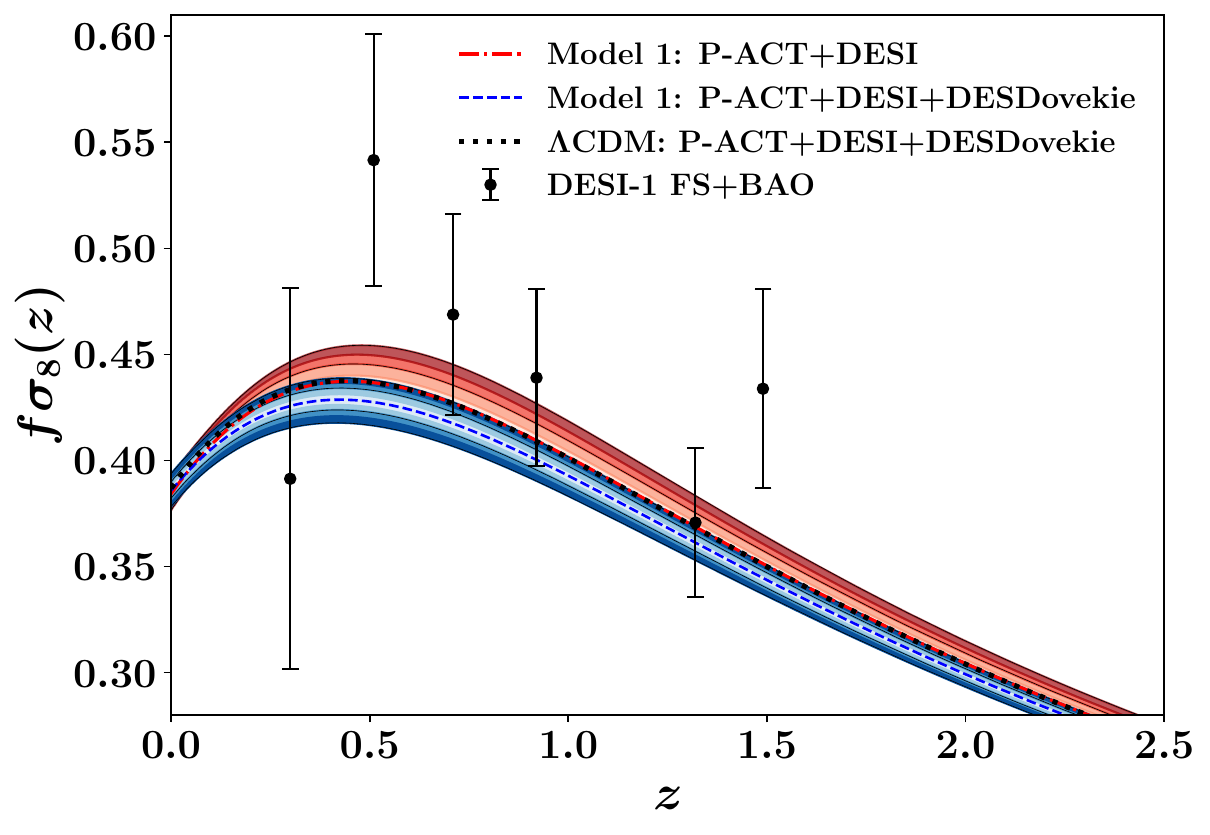}
    \includegraphics[width=0.325\linewidth]{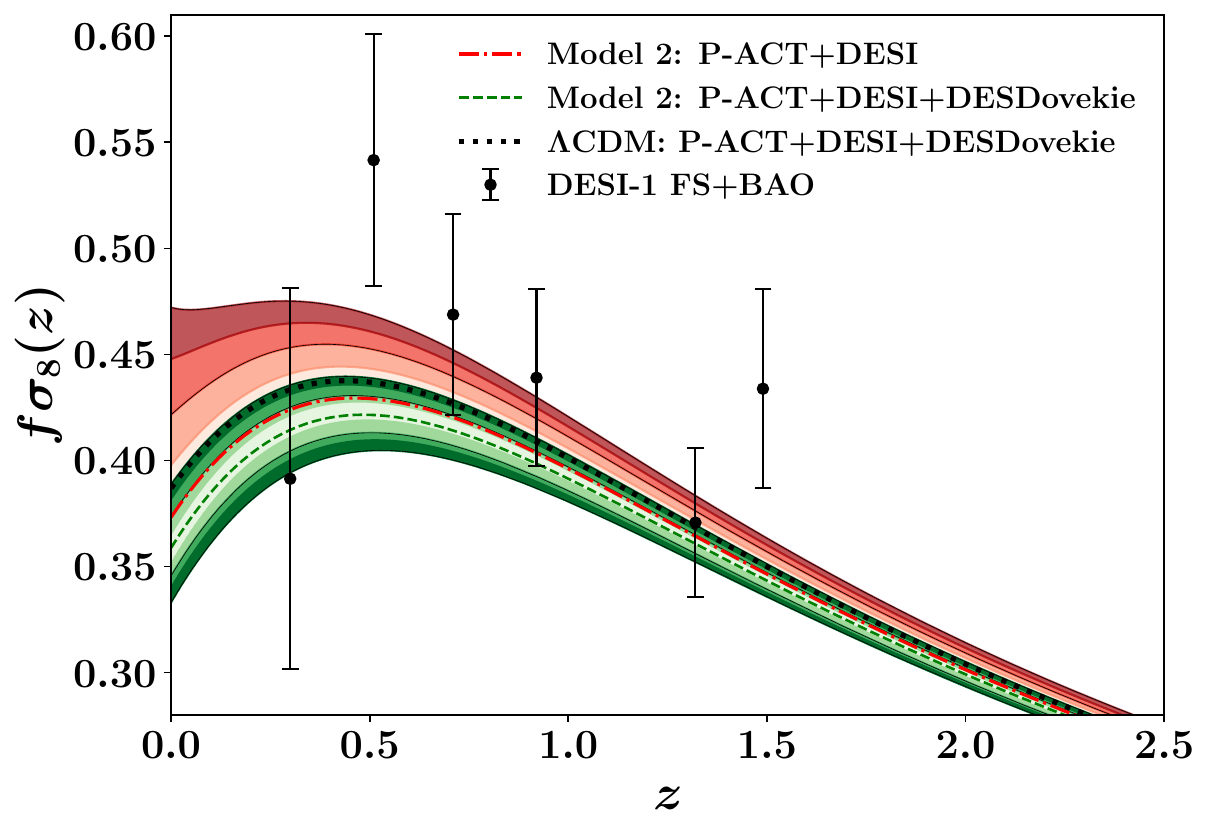}
    \includegraphics[width=0.325\linewidth]{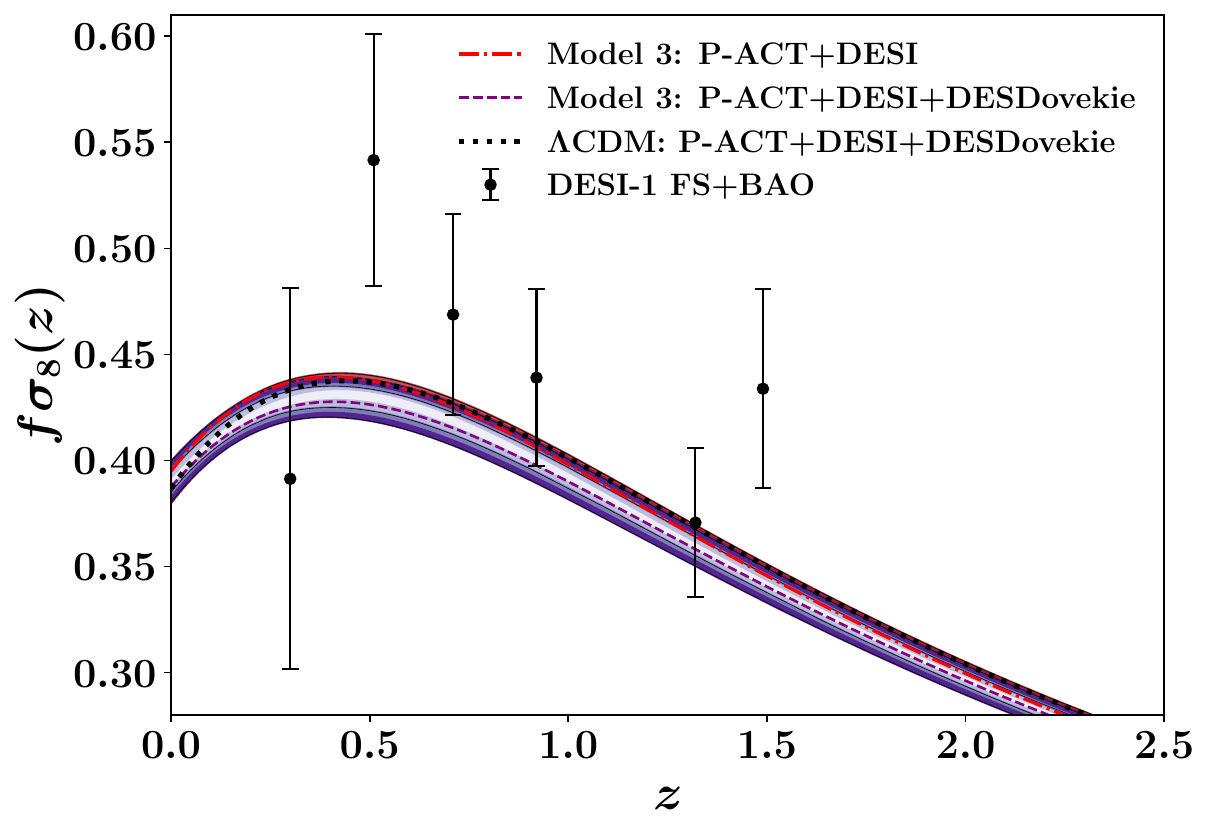}
    \caption{Reconstructed growth rate of structure $f\sigma_8(z)$ for the metastable DE models compared with DESI full-shape plus BAO measurements. Model 2 shows the largest growth-sector response because DE decay directly modifies the DM abundance and perturbations.}
    \label{fig:fsigma8z}
\end{figure*}

Figure~\ref{fig:qz} shows the deceleration parameter \cite{osti_4191853}, \begin{equation}
q(z) = -1+\frac{1+z}{H(z)}\frac{{\rm d}H(z)}{{\rm d}z}\, .
\end{equation} 
Positive values, $q(z)>0$, correspond to decelerated expansion, while negative values, $q(z)<0$, indicate accelerated expansion. All three metastable DE models show the expected transition from deceleration at higher redshift to acceleration at late times. The present-day best-fit values of $q_0$ are slightly less negative than in $\Lambda$CDM, indicating a mild slowing down of the present cosmic acceleration. This trend is qualitatively similar to trends seen in DESI results \cite{DESI:2025fii}. The transition redshift $z_t$, defined by $q(z_t)=0$, remains broadly consistent with $\Lambda$CDM expectation within $2\sigma$. The BBN+DESI+DES-Dovekie combination allows a broader range of late-time behaviour, while P-ACT information pulls the inferred $q(z)$ closer to the $\Lambda$CDM curve.

For the metastable DE models, the effective DE EoS is given by Eq.~\eqref{eq:weff_de}.  Positive values of $\Gamma/H_0$ correspond to quintessence-like DE behaviour, $w_{\rm DE}^{\rm eff}>-1$, while negative values correspond to phantom-like DE behaviour, $w_{\rm DE}^{\rm eff}<-1$. Figure~\ref{fig:wdez} shows the reconstructed evolution of $w_{\rm DE}(z)$. For the BBN+DESI+DES-Dovekie combination, Models 1 and 2 favour quintessence-like behaviour at low redshift at more than $2\sigma$ level. The P-ACT+DESI+DES-Dovekie constraints are tighter and remain closer to the $\Lambda$CDM limit. The agreement with $w_{\rm DE}=-1$ is model dependent: Model 2 remains consistent with $\Lambda$CDM within $2\sigma$, Model 1 lies close to the $2\sigma$ boundary, and Model 3 is marginally outside the $2\sigma$ region, although the deviation is small.  Thus, while the parent DE sector follows the same decay law in all three scenarios, the physical distinction between the models comes from the daughter component: DM in Model 2 and DR in Model 3. 

\begin{figure*}
    \centering
    \includegraphics[width=0.95\linewidth]{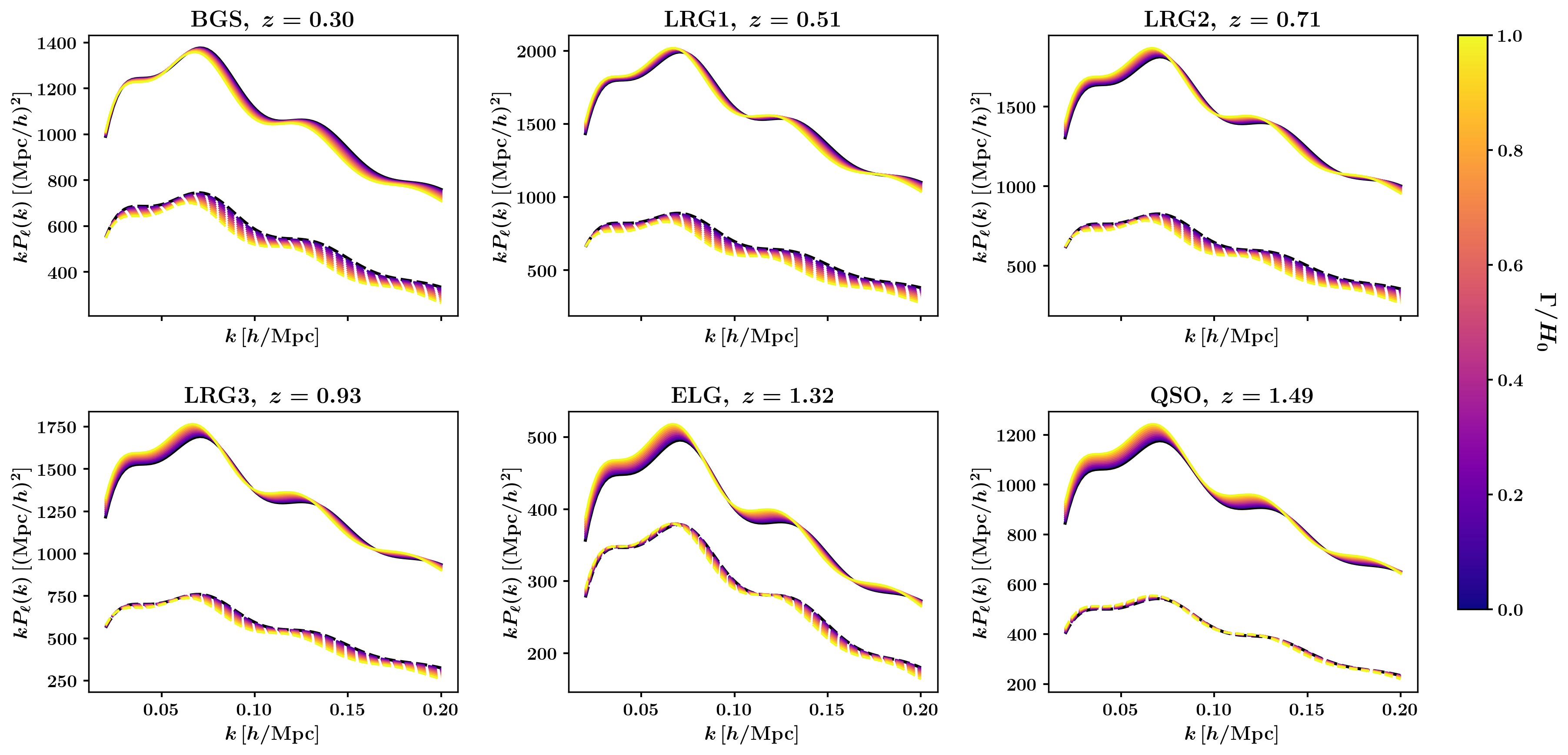}
    \caption{DESI DR1 full-shape power-spectrum multipoles for Model 1. The predictions show the response to different values of $\Gamma/H_0$ across the DESI tracer redshift bins. Since Model 1 mainly modifies the DE background, the changes in the multipoles remain mild.}
    \label{fig:fsbao_theo_m1}
\end{figure*}

Finally, we consider the growth of structure \cite{Huterer:2022dds} through $f\sigma_8(z)$, which combines the logarithmic growth rate of matter perturbations with the fluctuation amplitude on scales of $8\,h^{-1}{\rm Mpc}$, defined as $f\sigma_8(z)=f(z)\sigma_8(z)$, where $f(z)=\frac{d\ln \delta}{d\ln a}$ is the linear growth rate and $\delta(z)$ is normalized to $\delta(0)=1$. Since $\sigma_8(z)=\delta(z)\sigma_{8,0}$, this gives
\begin{equation}
    f\sigma_8(z) = - \,\sigma_{8,0}\, (1+z)\,\frac{{\rm d}\delta(z)}{{\rm d}z}.
\end{equation}
Figure~\ref{fig:fsigma8z} shows the inferred evolution of $f\sigma_8(z)$ for the three metastable DE models, together with the corresponding $\Lambda$CDM prediction and the DESI DR1 full-shape plus post-reconstruction BAO measurements \cite{DESI:2024jxi}, denoted as FS and extracted using the ShapeFit compression \cite{Brieden:2021edu}. 

The observable $f\sigma_8(z)$ is sensitive to both the background expansion and the evolution of matter perturbations. It therefore provides information complementary to geometric distance probes such as SN-Ia and BAO. While background quantities such as $\mathcal{O}m(z)$, $q(z)$, and $w_{\rm DE}(z)$ mainly probe the homogeneous expansion history, growth information helps distinguish the three metastable DE scenarios. This is because the evolution of structure is sensitive to whether the decay products cluster, as in the DM channel, or free-stream, as in the DR channel.

\begin{figure*}
    \centering
    \includegraphics[width=0.95\linewidth]{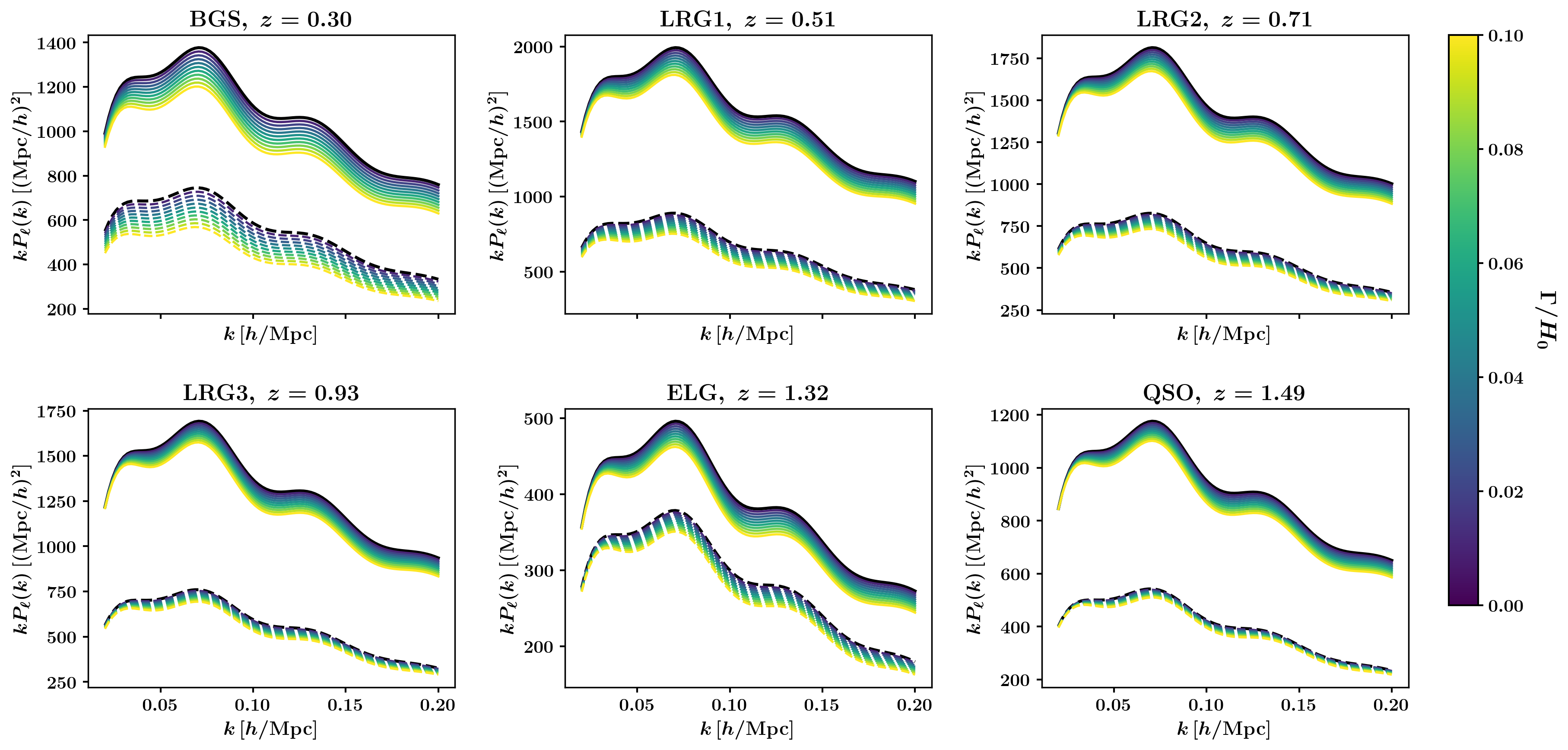}
    \caption{DESI DR1 full-shape power-spectrum multipoles for Model 2. The decay of DE into DM directly modifies the matter abundance and perturbations, leading to the strongest response in the clustering amplitude and scale dependence.}
    \label{fig:fsbao_theo_m2}
\end{figure*}

\begin{figure*}
    \centering
    \includegraphics[width=0.95\linewidth]{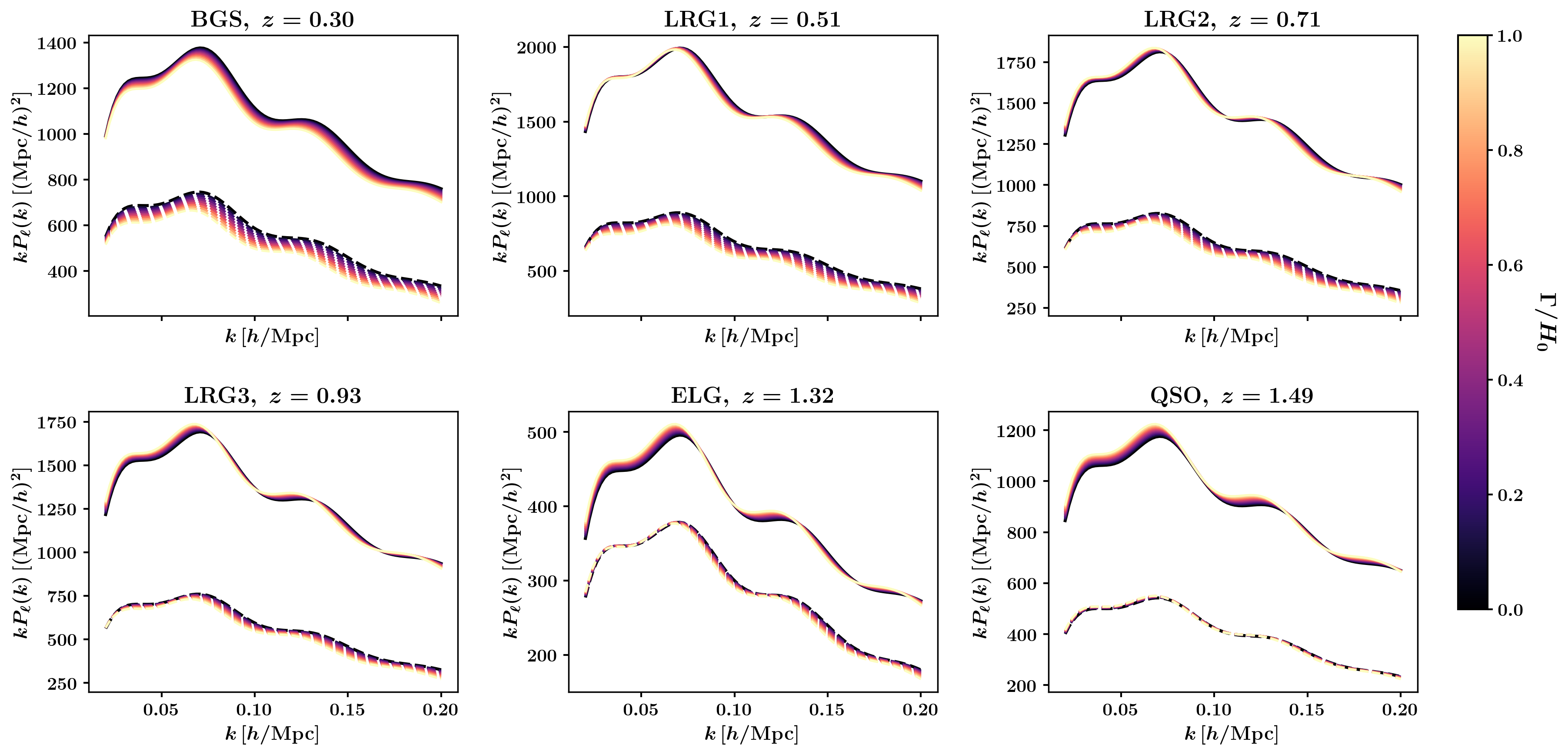}
    \caption{DESI DR1 full-shape power-spectrum multipoles for Model 3. The DR component produces a weaker clustering response than in Model 2.}
    \label{fig:fsbao_theo_m3}
\end{figure*}

\begin{table}
\centering
\caption{Cosmological priors adopted for the DESI DR1 FS analysis without CMB. For CMB+FS combinations, we use the CMB priors given in Table~\ref{tab:priors}.}
\label{tab:fs_priors}
\renewcommand{\arraystretch}{1.25}
\setlength{\tabcolsep}{15pt}
\begin{tabular}{c c}
\toprule
\textbf{Parameter} & \textbf{Prior} \\
\hline
${\boldmath \omega_c}$ & $\mathcal{U}(0.01,\,0.99)$ \\
${\boldmath \omega_b}$ & $\mathcal{N}(0.02218,\,0.00055^2)$ \\
${\boldmath H_0}$ & $\mathcal{U}(20,\,100)$ \\
${\boldmath \ln(10^{10}A_s)}$ & $\mathcal{U}(1.61,\,3.91)$ \\
${\boldmath n_s}$ & $\mathcal{N}(0.9649,\,0.042^2)$ \\
\midrule
$\Gamma/H_0$ & $\mathcal{U}(-1,\,2)$ \quad Models 1 and 2 \\
$\Gamma/H_0$ & $\mathcal{U}(0,\,2)$ \quad Model 3 \\
\bottomrule
\end{tabular}
\end{table}

\section{Impact of DESI DR1 Full Shape Measurements \label{sec:fs}}

\squeezetable
\begin{table*}
\centering
\caption{Marginalized constraints at 68\% CL for the baseline $\Lambda$CDM and metastable DE models with BBN+FS+$n_{s,10}$ (+DES-Dovekie) data combinations.}
\label{tab:fs_bbn_constraints}
\begin{minipage}[t]{0.85\textwidth}
\centering
\renewcommand{\arraystretch}{1.35}
\setlength{\tabcolsep}{7pt}
\resizebox{\textwidth}{!}{
\begin{tabular}{lcccc}
\toprule
\textbf{Parameter} & {\boldmath\textbf{$\Lambda$CDM}} & \textbf{Model 1} & \textbf{Model 2} & \textbf{Model 3} \\
\hline
\multicolumn{5}{c}{\textbf{BBN+FS+$n_{s,10}$}} \\
\hline
{\boldmath\textbf{$H_0$}} & $68.69\pm0.77$ & $71.8\pm2.2$ & $68.5\pm1.7$ & $66.3^{+1.8}_{-1.5}$ \\
{\boldmath\textbf{$\Omega_b h^2$}} & $0.02213\pm0.00054$ & $0.02217\pm0.00053$ & $0.02215\pm0.00052$ & $0.02215\pm0.00055$ \\
{\boldmath\textbf{$\Omega_c h^2$}} & $0.1171\pm0.0051$ & $0.1208\pm0.0057$ & $0.120^{+0.029}_{-0.023}$ & $0.1162\pm0.0052$ \\
{\boldmath\textbf{$\Gamma/H_0$}} & -- & $-0.43^{+0.26}_{-0.31}$ & $0.06\pm0.31$ & $<0.237$ \\
{\boldmath\textbf{$\Omega_m$}} & $0.2966\pm0.0094$ & $0.279^{+0.012}_{-0.016}$ & $0.308\pm0.068$ & $0.304\pm0.011$ \\
{\boldmath\textbf{$\Omega_{\rm DR}$}} & -- & -- & -- & $<0.0354$ \\
\hline
\multicolumn{5}{c}{\textbf{BBN+FS+$n_{s,10}$+DES-Dovekie}} \\
\hline
{\boldmath\textbf{$H_0$}} & $68.63\pm0.76$ & $67.6\pm1.0$ & $67.34\pm0.98$ & $66.2\pm1.3$ \\
{\boldmath\textbf{$\Omega_b h^2$}} & $0.02216\pm0.00056$ & $0.02214\pm0.00054$ & $0.02215\pm0.00055$ & $0.02215\pm0.00054$ \\
{\boldmath\textbf{$\Omega_c h^2$}} & $0.1217\pm0.0047$ & $0.1174^{+0.0051}_{-0.0057}$ & $0.1389\pm0.0093$ & $0.1167\pm0.0050$ \\
{\boldmath\textbf{$\Gamma/H_0$}} & -- & $0.16\pm0.11$ & $0.28\pm0.14$ & $0.23^{+0.10}_{-0.16}$ \\
{\boldmath\textbf{$\Omega_m$}} & $0.3067\pm0.0083$ & $0.3064\pm0.0085$ & $0.357\pm0.026$ & $0.3061\pm0.0081$ \\
{\boldmath\textbf{$\Omega_{\rm DR}$}} & -- & -- & -- & $0.034^{+0.015}_{-0.024}$ \\
\bottomrule
\end{tabular}
}
\end{minipage}
\end{table*}

\begin{figure*}
    \centering
    \includegraphics[height=0.21\textheight, width=0.325\linewidth]{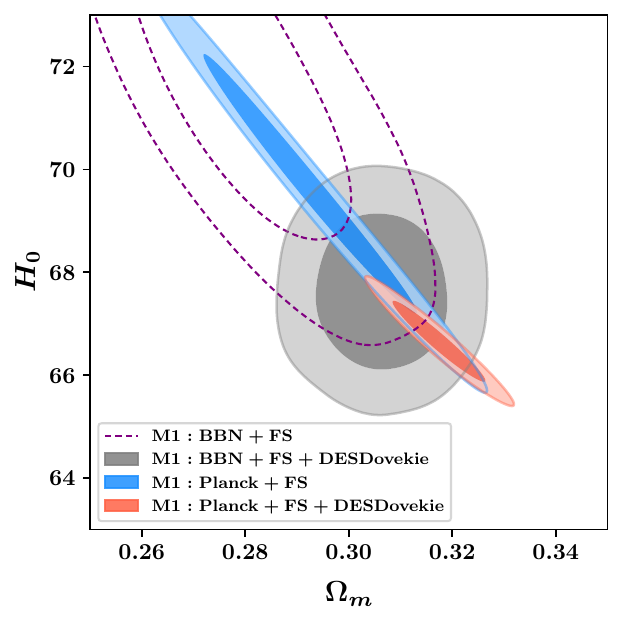}
    \includegraphics[height=0.21\textheight, width=0.325\linewidth]{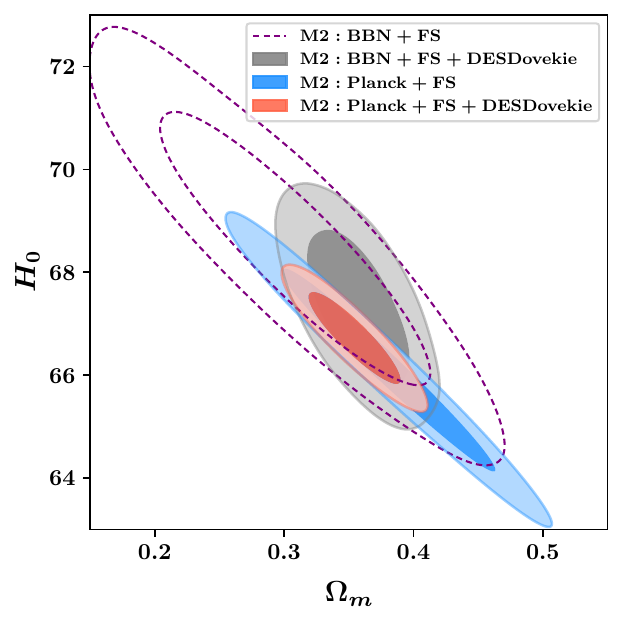}
    \includegraphics[height=0.21\textheight, width=0.325\linewidth]{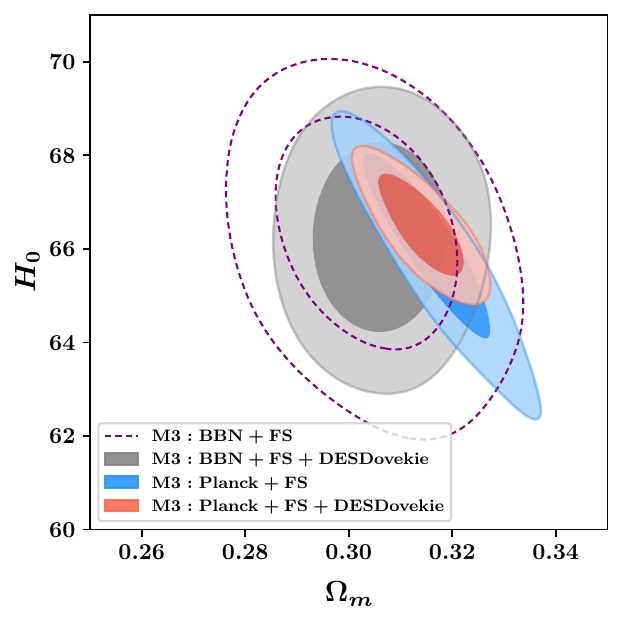}\\
    \includegraphics[height=0.21\textheight, width=0.325\linewidth]{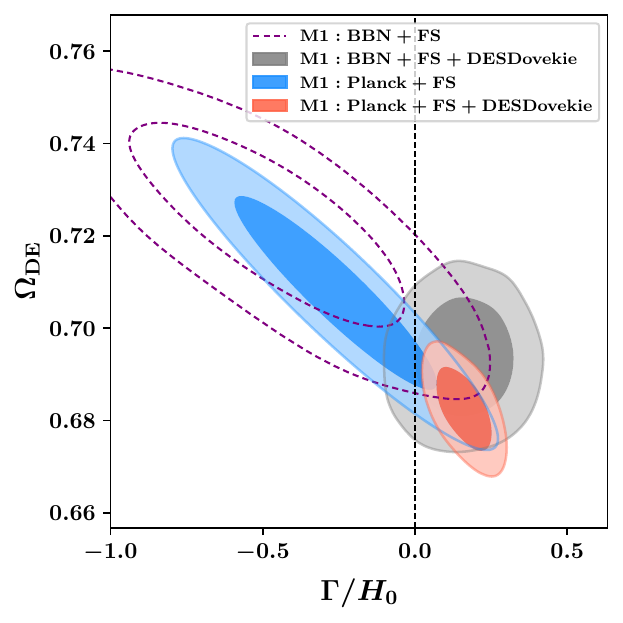}
    \includegraphics[height=0.21\textheight, width=0.325\linewidth]{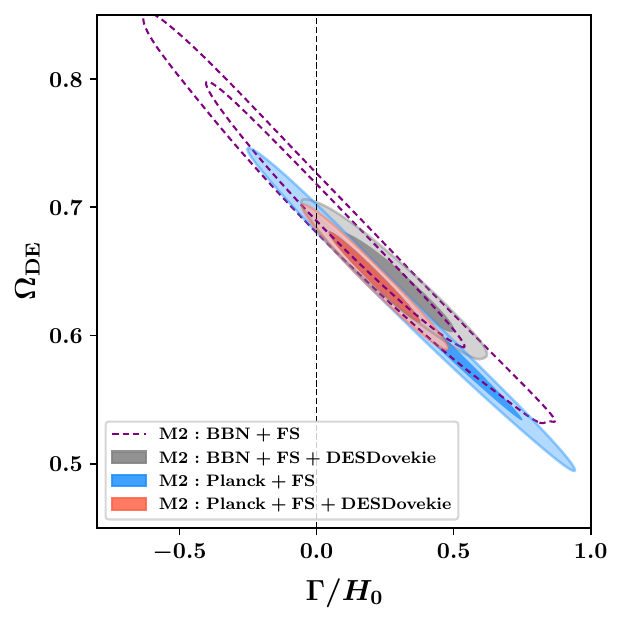}
    \includegraphics[height=0.21\textheight, width=0.325\linewidth]{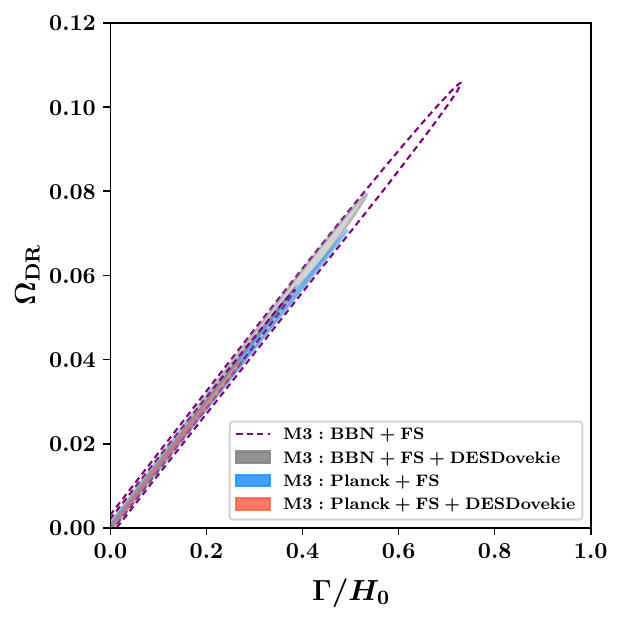}
    \caption{Impact of DESI DR1 FS measurements on the metastable DE 1,2 \& 3. The upper panels show the $\Omega_m-H_0$ constraints, and lower panels show the $\Gamma/H_0-\Omega_{\rm DE}$ or $\Gamma/H_0-\Omega_{\rm DR}$ constraints.}
    \label{fig:mcmc_fs}
\end{figure*}

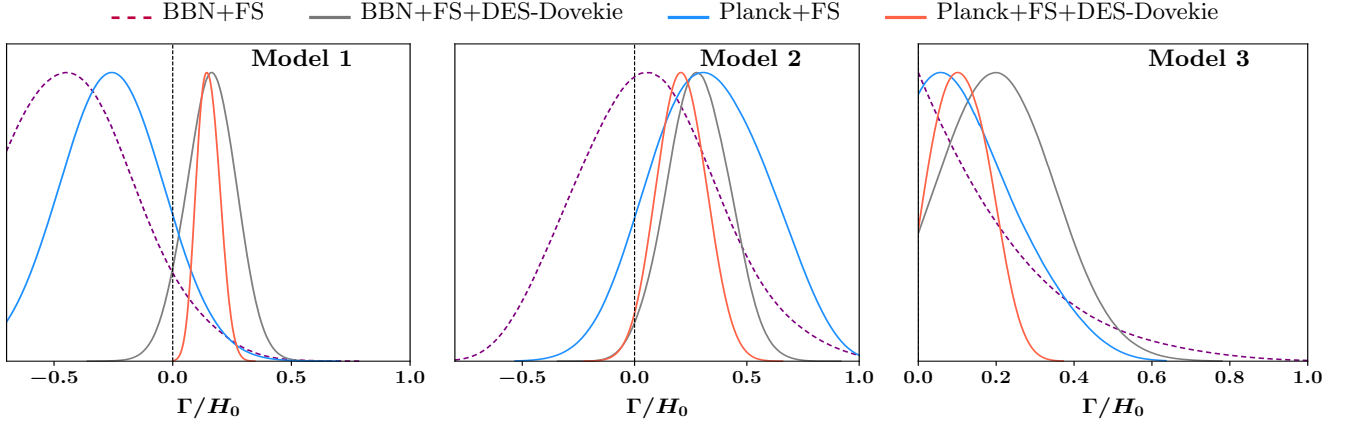
\begin{figure*}
    \centering

    \small
    \tikz{\draw[purple, dashed, line width=1.0pt] (0,0) -- (0.55,0);} BBN+FS
    \hspace{0.35cm}
    \tikz{\draw[mygray, solid, line width=1.2pt] (0,0) -- (0.55,0);} BBN+FS+DES-Dovekie
    \hspace{0.35cm}
    \tikz{\draw[dodgerblue, solid, line width=1.2pt] (0,0) -- (0.55,0);} Planck+FS
    \hspace{0.35cm}
    \tikz{\draw[tomato, solid, line width=1.2pt] (0,0) -- (0.55,0);} Planck+FS+DES-Dovekie

    \vspace{0.15cm}

    \begin{overpic}[height=0.22\textheight,width=0.325\linewidth]{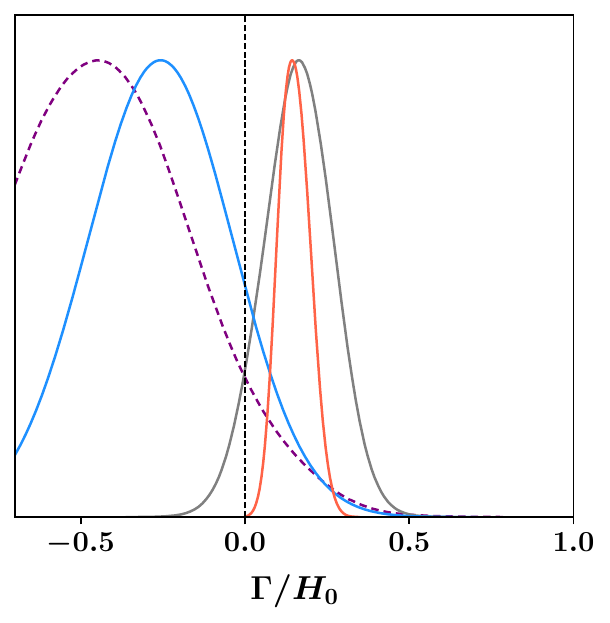}
        \put(58,82){\small\bfseries Model 1}
    \end{overpic}
    \begin{overpic}[height=0.22\textheight,width=0.325\linewidth]{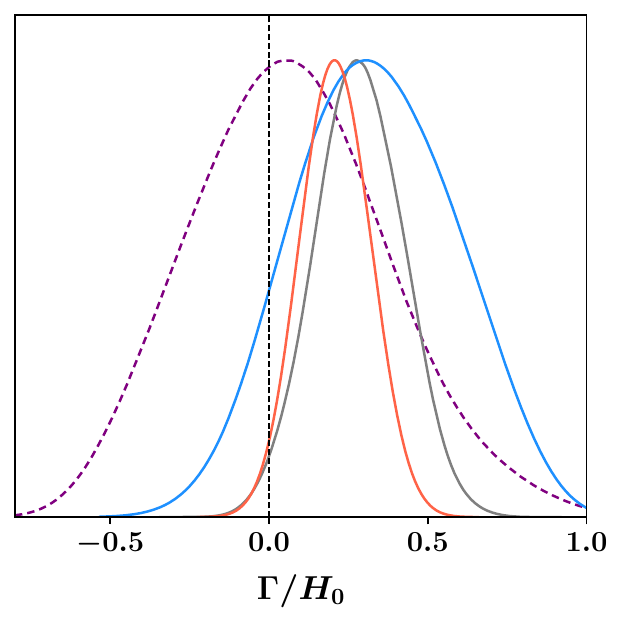}
        \put(58,82){\small\bfseries Model 2}
    \end{overpic}
    \begin{overpic}[height=0.22\textheight,width=0.325\linewidth]{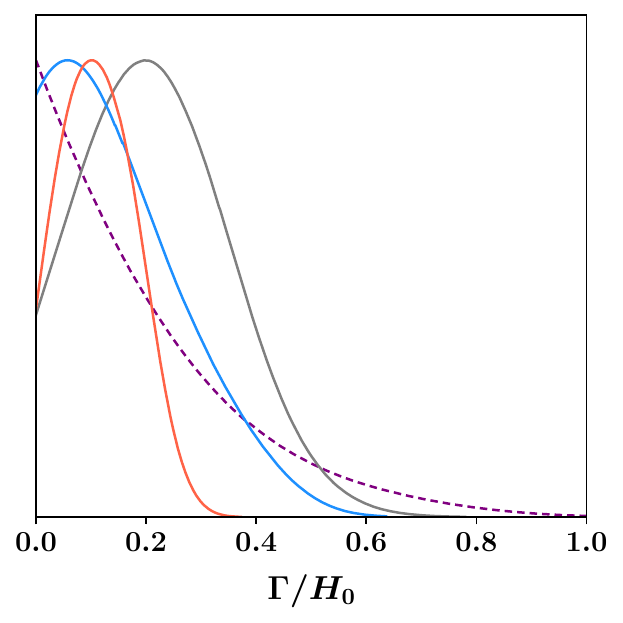}
        \put(58,82){\small\bfseries Model 3}
    \end{overpic}

    \caption{
    Marginalized 1D posterior distributions of $\Gamma/H_0$ for metastable dark energy Models 1--3 using the FS combinations. The vertical dashed line marks $\Gamma/H_0=0$.
    }
    \label{fig:mcmc2_fs}
\end{figure*}

\squeezetable
\begin{table*} 
\centering
\caption{Marginalized constraints at 68\% CL for the baseline $\Lambda$CDM and metastable DE models with Planck+FS (+DES-Dovekie) data combinations.}
\label{tab:fs_planck_constraints}
\begin{minipage}[t]{0.85\textwidth}
\centering
\renewcommand{\arraystretch}{1.35}
\setlength{\tabcolsep}{7pt}
\resizebox{\textwidth}{!}{
\begin{tabular}{lcccc}
\toprule
\textbf{Parameter} & {\boldmath\textbf{$\Lambda$CDM}} & \textbf{Model 1} & \textbf{Model 2} & \textbf{Model 3} \\
\hline
\multicolumn{5}{c}{\textbf{Planck+FS}} \\
\hline
{\boldmath\textbf{$H_0$}} & $67.79\pm0.37$ & $69.7\pm1.6$ & $66.2\pm1.2$ & $66.0^{+1.4}_{-1.1}$ \\
{\boldmath\textbf{$\Omega_b h^2$}} & $0.02225\pm0.00013$ & $0.02222\pm0.00013$ & $0.02228\pm0.00013$ & $0.02227\pm0.00013$ \\
{\boldmath\textbf{$\Omega_c h^2$}} & $0.11844\pm0.00081$ & $0.11894\pm0.00090$ & $0.142^{+0.019}_{-0.017}$ & $0.11818\pm0.00083$ \\
{\boldmath\textbf{$\Gamma/H_0$}} & -- & $-0.26\pm0.21$ & $0.33\pm0.24$ & $<0.202$ \\
{\boldmath\textbf{$\Omega_m$}} & $0.3076\pm0.0049$ & $0.292\pm0.013$ & $0.378\pm0.051$ & $0.3158^{+0.0074}_{-0.0086}$ \\
{\boldmath\textbf{$\Omega_{\rm DR}$}} & -- & -- & -- & $<0.0294$ \\
\hline
\multicolumn{5}{c}{\textbf{Planck+FS+DES-Dovekie}} \\
\hline
{\boldmath\textbf{$H_0$}} & $67.63\pm0.35$ & $66.65\pm0.49$ & $66.73\pm0.55$ & $66.49\pm0.66$ \\
{\boldmath\textbf{$\Omega_b h^2$}} & $0.02222\pm0.00013$ & $0.02226\pm0.00015$ & $0.02228\pm0.00013$ & $0.02226\pm0.00013$ \\
{\boldmath\textbf{$\Omega_c h^2$}} & $0.11878\pm0.00077$ & $0.11807\pm0.00075$ & $0.1346\pm0.0075$ & $0.11837\pm0.00085$ \\
{\boldmath\textbf{$\Gamma/H_0$}} & -- & $0.163\pm0.055$ & $0.21\pm0.10$ & $0.121^{+0.054}_{-0.086}$ \\
{\boldmath\textbf{$\Omega_m$}} & $0.3098\pm0.0047$ & $0.3174\pm0.0056$ & $0.354\pm0.022$ & $0.3142\pm0.0052$ \\
{\boldmath\textbf{$\Omega_{\rm DR}$}} & -- & -- & -- & $0.0176^{+0.0079}_{-0.013}$ \\
\bottomrule
\end{tabular}
}
\end{minipage}
\end{table*}

\begin{figure*}
    \centering
    \includegraphics[height=0.22\textheight, width=.325\linewidth]{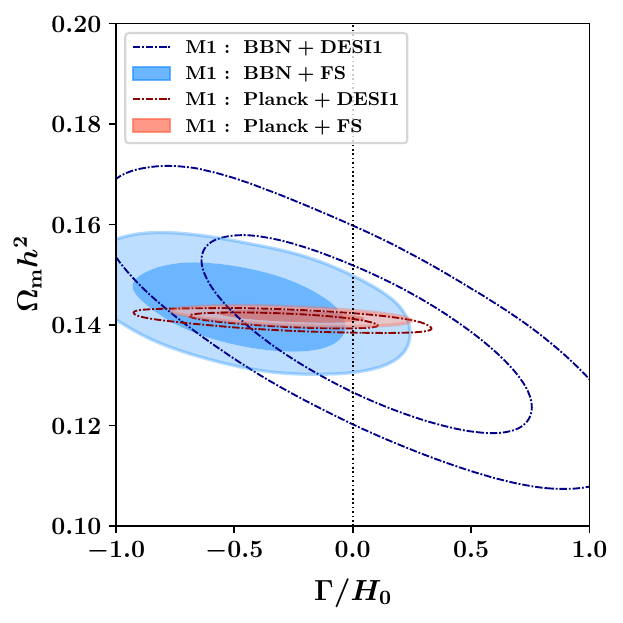}
    \includegraphics[height=0.22\textheight, width=.325\linewidth]{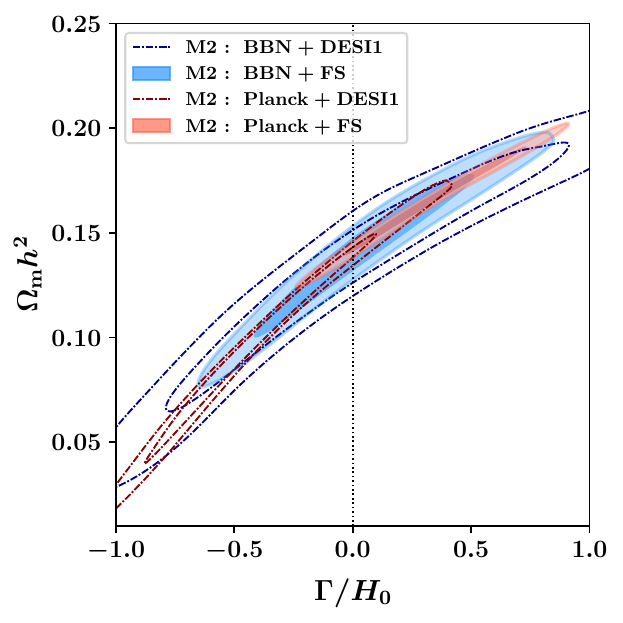}
    \includegraphics[height=0.22\textheight, width=.325\linewidth]{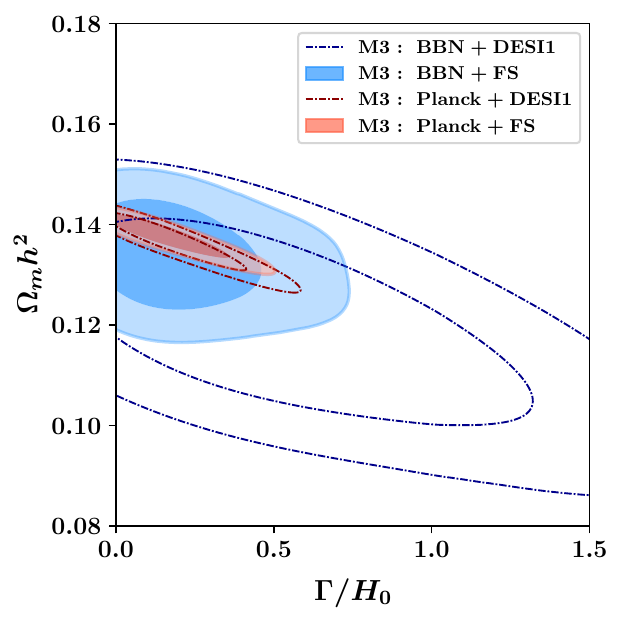}
    \caption{Comparison between DESI DR1 BAO and DESI DR1 FS+BAO constraints in the $\Omega_m h^2-\Gamma/H_0$ plane for metastable DE Models 1, 2 and 3.}
\label{fig:fs_vs_bao}
\end{figure*}

The reconstructed growth history discussed above motivates a closer look at galaxy clustering beyond geometric distance probes. While BAO and SN-Ia mainly constrain the background expansion, the broadband shape of the galaxy power spectrum is sensitive to the amplitude and scale dependence of structure growth. Full-shape measurements can therefore help distinguish the three metastable DE scenarios, since the decay affects perturbations differently in each model.

In this section, we illustrate the impact of the decay parameter on the DESI DR1 full-shape power-spectrum multipoles. We compare the theoretical predictions of the three models with the DESI DR1 full-shape measurements combined with post-reconstruction BAO constraints, hereafter denoted as FS. The predictions are computed using our modified \texttt{CLASS} implementation for the linear background and growth evolution. We interface this with \texttt{CosmoPrimo}\footnote{\url{https://github.com/cosmodesi/cosmoprimo.git}} through a modified \texttt{pyclass}\footnote{\url{https://github.com/adematti/pyclass.git}} engine, and pass the resulting linear power spectra, growth quantities, and background distances to \texttt{velocileptors}\footnote{\url{https://github.com/sfschen/velocileptors.git}} \cite{Maus:2024dzi}, following the DESI DR1 FS analysis \cite{DESI:2024jxi} framework.

The FS multipoles are evaluated for the DESI DR1 tracers, namely BGS, LRG, ELG, and QSO, at effective redshifts. Figures~\ref{fig:fsbao_theo_m1}, \ref{fig:fsbao_theo_m2}, and \ref{fig:fsbao_theo_m3} show the monopole and quadrupole predictions for Models 1-3. To isolate the effect of $\Gamma/H_0$, we fix all other cosmological and nuisance parameters to the best-fit Planck $\Lambda$CDM values. The colour variation therefore shows how the FS clustering signal responds to metastable DE dynamics.

Figure~\ref{fig:fsbao_theo_m1} shows that Model 1 produces only a mild response in the FS multipoles, since the decay mainly affects the DE background evolution and changes the clustering signal only through the expansion history and growth factor. In contrast, Fig.~\ref{fig:fsbao_theo_m2} shows a much stronger response for Model 2, where DE decays into non-baryonic DM. The variation with $\Gamma/H_0$ is visible in both the amplitude and scale dependence of the clustering signal, reflecting the direct modification of the DM abundance and matter perturbations. This makes DESI DR1 FS measurements particularly constraining for Model 2. Finally, Fig.~\ref{fig:fsbao_theo_m3} shows that Model 3 gives a weaker response, closer to Model 1, because the produced DR component remains subdominant at late times and affects the galaxy power spectrum mainly through the expansion history.

For the FS analysis, we vary the model parameters
$\left\{\omega_c,\, 
\omega_b,\, 
H_0,\, 
\ln(10^{10}A_s),\, 
n_s,\, 
\Gamma/H_0 
\right\}$, following the methodology of the DESI DR1 full-shape analysis. We fix $\tau_{\rm reio}$ to its Planck $\Lambda$CDM fiducial value. The baryon density is constrained by the BBN prior in Eq.~\eqref{eq:bbn} and $n_s$ is assigned the weak Gaussian prior $n_{s,10}$. The prior on $\Gamma/H_0$ is kept the same as in Table~\ref{tab:priors}. For the non-cosmological FS parameters, including bias, counterterm, stochastic, and nuisance parameters, we adopt the baseline DESI choices. The FS cosmological priors are summarized in Table~\ref{tab:fs_priors}. However, for the CMB+FS combinations, we instead adopt the priors listed in Table~\ref{tab:priors} and apply the same treatment for the nuisance parameters in the DESI DR1 FS likelihood.

Note that marginalized FS constraints can be sensitive to projection effects from nuisance-parameter marginalisation, as discussed in the DESI DR1 FS analysis \cite{DESI:2024jxi}. These effects are particularly relevant for extended models (e.g., $w_0w_a$CDM) that are not tightly constrained by FS or CMB+FS data alone. In our analysis, this caveat mainly concerns the BBN+FS, and, to a lesser extent, CMB+FS combinations. Nonetheless, once SN-Ia data are included, the qualitative conclusions are less sensitive to these effects.

Figure~\ref{fig:mcmc_fs} shows the impact of DESI DR1 FS measurements on the three metastable DE models. The upper panels show the $\Omega_m$--$H_0$ constraints, while the lower panels show the corresponding $\Gamma/H_0$--$\Omega_{\rm DE}$ planes for Models 1 and 2, and the $\Gamma/H_0$--$\Omega_{\rm DR}$ plane for Model 3. The marginalized one-dimensional posteriors of $\Gamma/H_0$ are shown in Fig.~\ref{fig:mcmc2_fs}. For Model 1, the FS combinations constrain the $\Gamma/H_0-\Omega_{\rm DE}$ degeneracy. The BBN+FS contours are broad, while adding DES-Dovekie tightens the constraints and shifts the posterior toward positive $\Gamma/H_0$. Also, Planck+FS extends toward negative $\Gamma/H_0$, but Planck+FS+DES-Dovekie again favours positive values and gives the tightest posterior. For Model 2, the impact is stronger: both the BBN+FS and Planck+FS constraints show a strong $\Gamma/H_0-\Omega_{\rm DE}$ anti-correlation, reflecting the direct effect of DE decay into DM on the matter abundance and perturbations. Adding DES-Dovekie shifts the posterior toward positive $\Gamma/H_0$, although the $\Gamma/H_0=0$ limit remains within the allowed $2\sigma$ region. For Model 3, the $\Gamma/H_0-\Omega_{\rm DR}$ plane shows the expected positive correlation, since larger decay rates produce larger DR abundances. The BBN+FS constraint is broad, while DES-Dovekie and Planck information restrict the allowed decay rate, with the strongest constraint obtained for Planck+FS+DES-Dovekie.

The marginalized constraints are summarized in Tables~\ref{tab:fs_bbn_constraints} and \ref{tab:fs_planck_constraints}. For the final Planck+FS+DES-Dovekie combination, Model 1 shows the strongest deviation, with $\Gamma/H_0=0$ outside the $2\sigma$ region (although within 3$\sigma$ CL), whereas Models 2 and 3 remain consistent with the $\Lambda$CDM limit within $2\sigma$. Overall, FS information provides a useful consistency test of whether the parameter shifts allowed by distance probes are compatible with growth and clustering measurements.

To make the impact of FS information more explicit, Fig.~\ref{fig:fs_vs_bao} compares DESI DR1 BAO-only \cite{DESI:2024mwx} vs FS constraints in the $\Gamma/H_0-\Omega_m h^2$ plane. Adding FS information reduces the allowed region in all three models. The tightening is moderate for Model 1, but stronger for Model 2, where the contours are substantially compressed along the $\Gamma/H_0-\Omega_m h^2$ degeneracy direction. Model 3 also shows a reduction in the allowed decay rate. This demonstrates the added constraining power of full-shape clustering and motivates extending the analysis to upcoming DESI DR2 FS measurements.

\squeezetable
\begin{table*}
\centering
\caption{Relative goodness of fit and DIC comparison for the metastable DE models with respect to $\Lambda$CDM. We define $\Delta\chi^2=\chi^2_{\rm model}-\chi^2_{\Lambda{\rm CDM}}$ and $\Delta{\rm DIC}={\rm DIC}_{\rm model}-{\rm DIC}_{\Lambda{\rm CDM}}$ within each fixed data combination. Negative values indicate preference for the corresponding metastable DE model over $\Lambda$CDM.}
\label{tab:delta_chi2_dic}
\renewcommand{\arraystretch}{1.25}
\setlength{\tabcolsep}{15pt}
\resizebox{\textwidth}{!}{
\begin{tabular}{lcccccc}
\toprule
\textbf{Data combination}
& \multicolumn{2}{c}{\textbf{M1}}
& \multicolumn{2}{c}{\textbf{M2}}
& \multicolumn{2}{c}{\textbf{M3}} \\
\cline{2-7}
& $\Delta\chi^2$ & $\Delta{\rm DIC}$
& $\Delta\chi^2$ & $\Delta{\rm DIC}$
& $\Delta\chi^2$ & $\Delta{\rm DIC}$ \\
\midrule
{BBN+DESI+DESDovekie} & $-12.38$ & $-8.13$ & $-12.55$ & $-8.42$ & $-12.48$ & $-6.85$ \\
{Planck+DESI+DESDovekie} & $-4.25$ & $-0.44$ & $-4.36$ & $-0.33$ & $-3.15$ & $-0.68$ \\
{P-ACT+DESI+DESDovekie} & $-8.89$ & $-3.76$ & $-10.09$ & $-3.65$ & $-9.32$ & $-3.89$ \\
\midrule
{BBN+FS+$n_{s,10}$+DESDovekie} & $-6.29$ & $-1.55$ & $-8.09$ & $-4.14$ & $-5.27$ & $-1.93$ \\
{Planck+FS+DESDovekie} & $-3.68$ & $-1.44$ & $-4.96$ & $-0.98$ & $-4.22$ & $-1.78$ \\
\bottomrule
\end{tabular}
}
\end{table*}

\section{Model Comparison}
\label{sec:model_comparison}

We compare the metastable DE models with $\Lambda$CDM using $\Delta\chi^2=\chi^2_{\rm model}-\chi^2_{\Lambda{\rm CDM}}$, evaluated separately for each fixed data combination. Negative values indicate an improved fit, while small $\vert\Delta\chi^2\vert$ values should be interpreted as statistically comparable fits rather than as decisive evidence for either model. As shown in Table~\ref{tab:delta_chi2_dic}, all three metastable DE scenarios improve the best-fit quality relative to $\Lambda$CDM for both the DESI DR2 BAO and DESI DR1 FS+BAO data-based combinations considered here. The improvements appear more pronounced for the BBN- and P-ACT-data combinations, while the Planck and FS data combinations show relatively modest gains.

The DIC comparison provides a complementary assessment by incorporating the posterior-averaged fit quality and effective model complexity. We define $\Delta{\rm DIC}={\rm DIC}_{\rm model}-{\rm DIC}_{\Lambda{\rm CDM}}$, with $\Delta{\rm DIC}<0$ favouring the metastable DE models. All combinations listed in Table~\ref{tab:delta_chi2_dic} yield negative $\Delta{\rm DIC}$ values, with a noticeable preference for the BBN+DESI+DES-Dovekie combination. The remaining DESI- and FS-based combinations indicate weaker to moderate improvements over $\Lambda$CDM.

\section{Summary \label{sec:summary}}

In this work, we revisited metastable dark energy models governed by a radioactive-like decay law. We considered three scenarios: Model 1, where the DE sector decays effectively as a background component; Model 2, where DE decays into non-baryonic DM and behaves as an interacting DM-DE scenario; and Model 3, where DE decays into DR. These models allow us to test whether current cosmological observations favour departures from the $\Lambda$CDM limit, $\Gamma/H_0=0$, and to identify whether any such departures arise predominantly from the background evolution of dark energy, the matter sector, or the radiation sector.

A key novelty of this work is the inclusion of the currently available DESI DR1 full-shape measurements in the analysis of metastable DE models. This extends the investigation beyond background probes and enables a direct test of the growth and clustering signatures associated with the different decay channels. Model 1 primarily modifies the background expansion and induces only mild changes in $f\sigma_8(z)$. Model 2 produces the strongest growth signature, since the decay directly alters both the DM abundance and its perturbations, while Model 3 generates a distinct but tightly constrained DR contribution. Full-shape information, therefore, helps discriminate among the metastable models.

The preference for a nonzero decay rate remains mild and dataset-dependent. BAO data alone are consistent with $\Gamma/H_0=0$, while adding SN-Ia data shifts the constraints toward positive $\Gamma/H_0$, corresponding to a decaying DE density and quintessence-like effective behaviour at low redshift. This shift reaches the $2\sigma$ level in some combinations, but does not provide decisive evidence for metastable dynamics. In CMB-informed combinations, the constraints tighten significantly, and the $\Lambda$CDM limit remains within, or close to, the allowed $2\sigma$ region for these models. The DESI DR1 full-shape information provides complementary constraints through the growth and clustering sector, particularly for Model 2. For the combined analysis, the $\Lambda$CDM limit lies within the $2\sigma$ CL for Models 2 and 3, and within $3\sigma$ for Model 1.

Overall, metastable DE remains a viable phenomenological model. The current formulation does not naturally realize phantom crossing for DE EoS, which may require a more general composite dark-sector framework, such as metastable DE embedded in braneworld cosmology \cite{Sahni:2002dx, Mishra:2025goj} or coupled with a negative cosmological constant/AdS-like vacuum \cite{Sen:2021wld, Mukherjee:2025myk} contribution. Future full-shape measurements, such as DESI DR2, can help improve the constraints on metastable models through more precise measurements of growth and clustering. Together with Euclid \cite{Euclid:2025pzh}, Rubin/LSST \cite{LSSTDarkEnergyScience:2012kar}, and improved low-redshift SN-Ia samples such as Zwicky Transient Facility (ZTF) \cite{Graham:2019qsw}, these will be crucial for testing whether the mild preference for positive $\Gamma/H_0$ persists with improved control of growth, clustering, calibration, and systematics.
 
\begin{acknowledgments}
{VS thanks the Anusandhan National Research
Foundation (ANRF), India, for the National Science
Chair Professorship, which provided partial funding for
this work.} The authors acknowledge the use of computational resources of the high-performance computing cluster \textit{Jindeok} at the Korea Astronomy and Space Science Institute.\\
\end{acknowledgments}

\appendix

\section{Impact of SN-Ia Compilations \label{app:snia}}

\squeezetable
\begin{table*}
\centering
\caption{Marginalized constraints at 68\% CL for the baseline $\Lambda$CDM and metastable DE models with BBN+DESI+Union3 and BBN+DESI+PantheonPlus data combinations. The last row reports $\Delta\chi^2$ for each fixed data combination.}
\label{tab:app_bbn}
\renewcommand{\arraystretch}{1.35}
\setlength{\tabcolsep}{7pt}
\resizebox{0.8\textwidth}{!}{
\begin{tabular}{lcccc}
\toprule
\textbf{Parameter} & {\boldmath\textbf{$\Lambda$CDM}} & \textbf{Model 1} & \textbf{Model 2} & \textbf{Model 3} \\
\hline
\multicolumn{5}{c}{\textbf{BBN+DESI+Union3}} \\
\hline
{\boldmath\textbf{$H_0$}} & $68.64\pm0.61$ & $65.2\pm1.4$ & $65.3\pm1.4$ & $62.4\pm2.0$ \\
{\boldmath\textbf{$\Omega_b h^2$}} & $0.02220\pm0.00056$ & $0.02218\pm0.00055$ & $0.02219\pm0.00056$ & $0.02223\pm0.00054$ \\
{\boldmath\textbf{$\Omega_c h^2$}} & $0.1206\pm0.0049$ & $0.1063\pm0.0066$ & $0.153^{+0.013}_{-0.010}$ & $0.1066\pm0.0071$ \\
{\boldmath\textbf{$\Gamma/H_0$}} & -- & $0.46\pm0.17$ & $0.62\pm0.24$ & $0.62^{+0.27}_{-0.31}$ \\
{\boldmath\textbf{$\Omega_m$}} & $0.3043\pm0.0084$ & $0.3040\pm0.0084$ & $0.413\pm0.041$ & $0.3049\pm0.0087$ \\
{\boldmath\textbf{$\Omega_{\rm DR}$}} & -- & -- & -- & $0.092^{+0.038}_{-0.044}$ \\
\hline
\multicolumn{5}{c}{\textbf{BBN+DESI+PantheonPlus}} \\
\hline
{\boldmath\textbf{$H_0$}} & $68.65\pm0.58$ & $66.6\pm1.1$ & $66.6\pm1.1$ & $64.8\pm1.4$ \\
{\boldmath\textbf{$\Omega_b h^2$}} & $0.02221\pm0.00053$ & $0.02218\pm0.00053$ & $0.02218\pm0.00055$ & $0.02217\pm0.00055$ \\
{\boldmath\textbf{$\Omega_c h^2$}} & $0.1206\pm0.0045$ & $0.1109^{+0.0058}_{-0.0065}$ & $0.1388\pm0.0093$ & $0.1110\pm0.0063$ \\
{\boldmath\textbf{$\Gamma/H_0$}} & -- & $0.27\pm0.12$ & $0.35\pm0.17$ & $0.35^{+0.16}_{-0.21}$ \\
{\boldmath\textbf{$\Omega_m$}} & $0.3043\pm0.0078$ & $0.3015\pm0.0080$ & $0.365\pm0.030$ & $0.3019\pm0.0081$ \\
{\boldmath\textbf{$\Omega_{\rm DR}$}} & -- & -- & -- & $0.052^{+0.024}_{-0.031}$ \\
\bottomrule
\end{tabular}
}
\end{table*}

\begin{table*}
\centering
\caption{Marginalized constraints at 68\% CL for the baseline $\Lambda$CDM and metastable DE models with Planck+DESI+Union3 and Planck+DESI+PantheonPlus data combinations.}
\label{tab:app_plc}
\renewcommand{\arraystretch}{1.35}
\setlength{\tabcolsep}{7pt}
\resizebox{0.8\textwidth}{!}{
\begin{tabular}{lcccc}
\toprule
\textbf{Parameter} & {\boldmath\textbf{$\Lambda$CDM}} & \textbf{Model 1} & \textbf{Model 2} & \textbf{Model 3} \\
\hline
\multicolumn{5}{c}{\textbf{Planck+DESI+Union3}} \\
\hline
{\boldmath\textbf{$\ln(10^{10}A_s)$}} & $3.047\pm0.014$ & $3.050^{+0.014}_{-0.015}$ & $3.050\pm0.014$ & $3.051\pm0.015$ \\
{\boldmath\textbf{$n_s$}} & $0.9676\pm0.0033$ & $0.9688\pm0.0036$ & $0.9687\pm0.0034$ & $0.9693\pm0.0035$ \\
{\boldmath\textbf{$H_0$}} & $68.11\pm0.28$ & $67.35\pm0.77$ & $67.39\pm0.74$ & $66.0^{+1.1}_{-0.91}$ \\
{\boldmath\textbf{$\Omega_b h^2$}} & $0.02231\pm0.00013$ & $0.02234^{+0.00013}_{-0.00012}$ & $0.02233\pm0.00013$ & $0.02235\pm0.00012$ \\
{\boldmath\textbf{$\Omega_c h^2$}} & $0.11775\pm0.00062$ & $0.11733\pm0.00074$ & $0.130\pm0.011$ & $0.11717\pm0.00071$ \\
{\boldmath\textbf{$\tau_{\rm reio}$}} & $0.0591\pm0.0069$ & $0.0606\pm0.0074$ & $0.0606\pm0.0074$ & $0.0614^{+0.0071}_{-0.0079}$ \\
{\boldmath\textbf{$\Gamma/H_0$}} & -- & $0.12\pm0.11$ & $0.15\pm0.15$ & $<0.259$ \\
{\boldmath\textbf{$\Omega_m$}} & $0.3034\pm0.0036$ & $0.3095\pm0.0068$ & $0.337\pm0.032$ & $0.3115\pm0.0062$ \\
{\boldmath\textbf{$\Omega_{\rm DR}$}} & -- & -- & -- & $0.0298^{+0.0093}_{-0.028}$ \\
\hline
\multicolumn{5}{c}{\textbf{Planck+DESI+PantheonPlus}} \\
\hline
{\boldmath\textbf{$\ln(10^{10}A_s)$}} & $3.047\pm0.014$ & $3.049\pm0.014$ & $3.049^{+0.013}_{-0.015}$ & $3.050\pm0.014$ \\
{\boldmath\textbf{$n_s$}} & $0.9679\pm0.0035$ & $0.9686\pm0.0036$ & $0.9685\pm0.0034$ & $0.9688\pm0.0034$ \\
{\boldmath\textbf{$H_0$}} & $68.11\pm0.29$ & $67.57\pm0.59$ & $67.61\pm0.57$ & $66.57^{+0.89}_{-0.73}$ \\
{\boldmath\textbf{$\Omega_b h^2$}} & $0.02231\pm0.00012$ & $0.02233\pm0.00012$ & $0.02233\pm0.00013$ & $0.02234\pm0.00012$ \\
{\boldmath\textbf{$\Omega_c h^2$}} & $0.11777\pm0.00062$ & $0.11738\pm0.00071$ & $0.1262^{+0.0092}_{-0.0077}$ & $0.11725\pm0.00067$ \\
{\boldmath\textbf{$\tau_{\rm reio}$}} & $0.0589\pm0.0071$ & $0.0604^{+0.0067}_{-0.0076}$ & $0.0601^{+0.0063}_{-0.0076}$ & $0.0607\pm0.0073$ \\
{\boldmath\textbf{$\Gamma/H_0$}} & -- & $0.088\pm0.086$ & $0.11\pm0.11$ & $<0.184$ \\
{\boldmath\textbf{$\Omega_m$}} & $0.3034\pm0.0037$ & $0.3075\pm0.0054$ & $0.327\pm0.024$ & $0.3088^{+0.0048}_{-0.0054}$ \\
{\boldmath\textbf{$\Omega_{\rm DR}$}} & -- & -- & -- & $<0.0273$ \\
\bottomrule
\end{tabular}
}
\end{table*}

\begin{table*}
\centering
\caption{Marginalized constraints at 68\% CL for the baseline $\Lambda$CDM and metastable DE models with P-ACT+DESI+Union3 and P-ACT+DESI+PantheonPlus data combinations.}
\label{tab:app_pact}
\renewcommand{\arraystretch}{1.35}
\setlength{\tabcolsep}{7pt}
\resizebox{0.8\textwidth}{!}{
\begin{tabular}{lcccc}
\toprule
\textbf{Parameter} & {\boldmath\textbf{$\Lambda$CDM}} & \textbf{Model 1} & \textbf{Model 2} & \textbf{Model 3} \\
\hline
\multicolumn{5}{c}{\textbf{P-ACT+DESI+Union3}} \\
\hline
{\boldmath\textbf{$\ln(10^{10}A_s)$}} & $3.062\pm0.011$ & $3.066^{+0.011}_{-0.012}$ & $3.066\pm0.012$ & $3.067\pm0.011$ \\
{\boldmath\textbf{$n_s$}} & $0.9747\pm0.0030$ & $0.9764\pm0.0032$ & $0.9757\pm0.0031$ & $0.9763\pm0.0032$ \\
{\boldmath\textbf{$H_0$}} & $68.40\pm0.27$ & $67.24\pm0.75$ & $67.48\pm0.63$ & $65.7^{+1.2}_{-1.1}$ \\
{\boldmath\textbf{$\Omega_b h^2$}} & $0.02255\pm0.00010$ & $0.02258\pm0.00010$ & $0.02257\pm0.00010$ & $0.02259\pm0.00010$ \\
{\boldmath\textbf{$\Omega_c h^2$}} & $0.11741\pm0.00065$ & $0.11665\pm0.00082$ & $0.133^{+0.011}_{-0.0092}$ & $0.11657\pm0.00076$ \\
{\boldmath\textbf{$\tau_{\rm reio}$}} & $0.0611^{+0.0056}_{-0.0067}$ & $0.0620^{+0.0055}_{-0.0071}$ & $0.0622\pm0.0070$ & $0.0623\pm0.0064$ \\
{\boldmath\textbf{$\Gamma/H_0$}} & -- & $0.18\pm0.11$ & $0.20\pm0.12$ & $0.27^{+0.12}_{-0.19}$ \\
{\boldmath\textbf{$\Omega_m$}} & $0.3005\pm0.0036$ & $0.3095\pm0.0066$ & $0.343^{+0.030}_{-0.027}$ & $0.3107\pm0.0067$ \\
{\boldmath\textbf{$\Omega_{\rm DR}$}} & -- & -- & -- & $0.039^{+0.018}_{-0.027}$ \\
\hline
\multicolumn{5}{c}{\textbf{P-ACT+DESI+PantheonPlus}} \\
\hline
{\boldmath\textbf{$\ln(10^{10}A_s)$}} & $3.063\pm0.011$ & $3.065^{+0.010}_{-0.012}$ & $3.065\pm0.011$ & $3.066\pm0.011$ \\
{\boldmath\textbf{$n_s$}} & $0.9748\pm0.0030$ & $0.9762\pm0.0031$ & $0.9757\pm0.0030$ & $0.9762\pm0.0031$ \\
{\boldmath\textbf{$H_0$}} & $68.41\pm0.28$ & $67.60\pm0.59$ & $67.67\pm0.55$ & $66.45\pm0.84$ \\
{\boldmath\textbf{$\Omega_b h^2$}} & $0.02255\pm0.00010$ & $0.02257\pm0.00011$ & $0.02256\pm0.00011$ & $0.02258\pm0.00011$ \\
{\boldmath\textbf{$\Omega_c h^2$}} & $0.11738\pm0.00067$ & $0.11680\pm0.00078$ & $0.1299\pm0.0082$ & $0.11672\pm0.00077$ \\
{\boldmath\textbf{$\tau_{\rm reio}$}} & $0.0616^{+0.0057}_{-0.0069}$ & $0.0616^{+0.0056}_{-0.0069}$ & $0.0622^{+0.0063}_{-0.0071}$ & $0.0619^{+0.0064}_{-0.0072}$ \\
{\boldmath\textbf{$\Gamma/H_0$}} & -- & $0.131\pm0.087$ & $0.16\pm0.11$ & $0.188^{+0.071}_{-0.16}$ \\
{\boldmath\textbf{$\Omega_m$}} & $0.3004\pm0.0037$ & $0.3065\pm0.0054$ & $0.335\pm0.023$ & $0.3073\pm0.0052$ \\
{\boldmath\textbf{$\Omega_{\rm DR}$}} & -- & -- & -- & $0.028^{+0.011}_{-0.023}$ \\
\bottomrule
\end{tabular}
}
\end{table*}

We summarize the corresponding constraints obtained with Union3 and PantheonPlus, in order to show how the results depend on the choice of the SN-Ia data set.


\bibliography{references}

\end{document}